\documentstyle[onecolumn]{mn}

\begin{document}

\title[Synchrotron emission from relativistic jets]
{Synchrotron emission from relativistic parsec--scale jets.}

\author[A. R. Beresnyak, Ya. N. Istomin and V. I. Pariev]
{A. R. Beresnyak$^1$\thanks{E-mail: beres@td.lpi.ac.ru},
Ya. N. Istomin$^1$\thanks{E-mail: istomin@td.lpi.ac.ru}
and V. I. Pariev$^{1,2}$\\
$^1$P.~N.~Lebedev Physical Institute,
Leninsky Prospect 53, Moscow 117924, Russia\\
$^2$Steward Observatory, University of Arizona, 933 North Cherry
Avenue, Tucson, AZ 85721, USA}

\maketitle

\begin{abstract}
We developed theory of particle acceleration inside the relativistic rotating
electron-positron force--free jet with spiral magnetic field. We considered
perturbation of stationary magnetic field structure and found that
acceleration takes place in the regions where the Alfv\'en resonant condition
with the eigenmodes in the jet is fulfilled, i.e. where the local Alfv\'en
speed is equal to the phase speed of an eigenmode.
Acceleration process and synchrotron losses combined together form 
power law energy
spectrum of ultrarelativistic electrons and positrons with index between 2 and
3 depending upon initial energy of injected particles.
Synchrotron emission
of these electrons and positrons in spiral magnetic field of rotating
force--free jet has been calculated. Polarization properties of the
radiation has been obtained and compared with
existing VLBI polarization measurements
of parsec--scale jets in BL~Lac sources and quasars.
Our results give natural explanation of observed
bimodality in alignment between electric field vector and jet axis.
Degree of polarization and velocity of observed proper motion of bright knots
depend upon angular rotational velocity of the jet. Thus, comparing them
to each other, we can estimate angular rotational velocity in jets.
These results indicate that the fact, that generally in BL~Lac objects
electric field vector is oriented parallel to the jet axis while in
quasars perpendicular to the jet axis, may be due to intrinsically larger
angular rotational velocity and large winding of magnetic field in BL~Lac
jets than in quasar jets.
\end{abstract}

\section{Introduction}
The nature of the Active Galactic Nuclei (AGN) is still unclear. 
The most common
point of view is that there is a supermassive black hole in the centre of
active galaxy with mass approximately $10^8 \div 10^9 M_\odot$~(Rees, 1984).
Accretion on the black hole leads to the formation of powerful turbulent
processes in accretion discs which effectively heat the plasma and
generate magnetic field
of the order of $10^4 G$. Ejection of plasma from discs with
frozen magnetic field and emission  lead to the formation of
collimated streams
which transfer energy to large distances, up to $10^5$ pc.
Besides, the processes of generation of electron--positron plasma
may occur in the magnetospheres
of the black holes as well as in the vicinity of accretion discs.
Particles are created in collisions
of high--energy photons produced by inverse Compton
scattering of ultraviolet photons
emitted from disk with fast particles accelerated by magnetic
reconnection within disk
or  within magnetospheres of the rotating black hole, where equilibrium charge
density changes its sign (see, for example Beskin, Istomin~\& Pariev, 1992b).
 Electron--positron plasma forms pinched streams (jets) which have radii of the
order of one parsec. The energy density of electromagnetic field in these jets
greatly exceeds the energy density of electrons and positrons
taking into account
their rest energy. Rotation of the black hole and the accretion disc is
transmitted
to the jet due to magnetic coupling, where strong
radial electric field $E_r$ arises. Drift in this field and in the longitudinal
magnetic field $B_z$ frozen into plasma leads to differential rotation.
Electric current transferred by the jet produces  azimuthal
magnetic field $B_\phi$,
and magnetic field lines are twisted into spirals.
Energy in the jet is transferred
by the Pointing vector which is proportional to the product of radial electric
field and azimuthal magnetic field. Therefore, there is no problem of
energy transfer along the jet on large distances as there is in the case
when the main energy is in the hydrodynamic motion.

 The most amazing facts of the jets from AGN are their high collimation
(the ratio of the length to the diameter is a few orders of magnitude),
stability and apparent superluminal velocities of distinct knots
(Begelman, Blandford \& Rees, 1984).
The hydrodynamic and MHD stability of jets was investigated
in many works (Blandford \& Pringle,
1976; Torricelli--Ciamponi \& Petrini, 1990; Appl, Camenzind, 1992).
As to the stability of the electron--positron jets, rotating and moving
with relativistic speed, it was shown in papers by Istomin \& Pariev, 1994 and
Istomin \& Pariev, 1996  that they are stable with respect to
axisymmetric as well as spiral perturbations. The physical reason for
stability is the shear of the magnetic field, which in the case of small
hydrodynamic pressure of plasma stabilizes small oscillations.
Perturbations do not increase with time (Im$\omega\equiv0$) or have small
damping decrement ($\mbox{Im}\omega\approx10^{-2}\mbox{Re}\omega$) because of
the
resonance with Alfv\'en waves $\omega^\prime=k_\|^\prime c$.
In electron--positron plasma, when the specific energy density and pressure
are much less than the
energy density of magnetic field, Alfv\'en velocity is equal to the speed
of light in vacuum $c$ ($\omega^\prime$ and $k^\prime$ are the frequency and
the wave vector in the plasma rest frame). Resonance condition is fulfilled
on the specific magnetic surface and, in the case of cylindrical jet, at
the definite distance from the axis. In the vicinity of that surface magnetic
and electric fields of the wave reach large magnitudes. The particles of the
jet are accelerated by the electric field here, which leads
to the absorption of
the energy of perturbation.
Thus, the stability of the jet is connected directly with the
production of energetic particles in the current.
This resolves the problem of
{\it in situ} acceleration of electrons and positrons
producing
synchrotron emission from knots observed along jets.
In fact, as it is known (see Begelman at al., 1984),
energetic particles accelerated in central region can not
penetrate in the jet, they
must loose their transverse momentum by synchrotron radiation
already in the basement of
the jet. Characteristic time of synchrotron losses  is given by
 $$ \tau_s=\frac34\frac{c\gamma_\|}{r_e\omega_{B0}^2}. $$
Here $\gamma_\|$ is the longitudinal Lorentz--factor of the particle,
$\gamma_\|=(1-v_\|^2/c^2)^{-1/2}$; $r_e$ is the classical
radius of the electron $r_e=e^2/mc^2$; $\omega_{B0}=eB/mc$ is the 
electron cyclotron
frequency. Given  $\gamma_\|=10$, $B=10^2G$ radiation time $\tau_s$
is approximately $3\cdot10^5 \mbox{sec}$ whereas the typical time of the flyby
through the region of strong fields $\tau_f\approx \ell/c\approx 10^6 
\mbox{sec}$.
Here $\ell$ is the diameter of the jet, $\ell\approx 3\cdot10^{16}\mbox{cm}$.

In this paper we assume that the energy transferred along the jet, is mainly
the electromagnetic energy which can be transmitted for long distances along
the jets as along the wires. In this case it is very possible that
part of that energy is in waves propagating along the jet, which are
eigenmodes of cylindrical beam. The source of the wave motion is
nonstationary processes in the magnetospheres of black hole and accretion disc.
Short time variability on scales from days to months is actually observed
in AGN~(Mushotzky, Done \& Pound, 1993; Witzel et al., 1993).
Such a picture also gives a natural explanation of the
superluminal motions but requires neither special orientation
of the jet nor relativistic jet with high speed ($\beta=u/c>\sqrt2/2$).
As it was shown in papers by Istomin~\& Pariev, 1994 and Istomin~\& Pariev,
1996 there exist so called standing modes among
eigenmodes in jets ($v_{group}\equiv0$) which are not propagate with finite
velocity along jet but are only subject to diffuse spreading and,
therefore, their amplitude is maximal.
Phase velocity of these modes is greater than the speed of light.
Crests of the wave move along the jet  with superluminal velocity causing
acceleration of the particles on the Alfv\'en surface. These regions can be
the objects which are seen
as bright knots with typical sizes of the order of the wavelength of the
standing wave which is about the radius of the jet.

The purpose of our paper is the consideration of the acceleration processes
near the Alfv\'en resonant surface, generation of the spectra of accelerated
particles and calculation of the polarization of synchrotron radiation
generated by these particles.

\section{Acceleration of particles}

The equations of motion near the Alfv\'en surface are rather complicated.
Nevertheless, we can use drift equations~(Sivukhin, 1965) because
the Larmor radius
of relativistic electrons and positrons $r_c={\cal E}/eB\approx10^7\mbox{cm}$
(if we take $B\approx 10^{-3}$ and $\gamma\approx 10$)
is much less then radius of the jet and the width of resonant surface $r_0$
(see expression (31*)).
\def\e{{\cal E}}
$$ \frac{d{\bf r}}{dt}=v_\|\frac{{\bf B}}B+\frac c{B^2}[{\bf E\times B}]+
                   \frac{\e v_\|^2}{ecB^4}[{\bf B\times(B\nabla)B}]+
                   \frac{\e v_\perp^2}{2ecB^3}[{\bf B\times\nabla}B];$$
$$ \frac{d\e}{dt}=e{\bf E}\frac{d{\bf r}}{dt}+
\frac{\e v_\perp^2}{2c^2B}\frac{\partial B}{\partial t}; \eqno (1) $$
$$ \frac{dJ_\perp}{dt}=0.$$
Here the electric field is assumed to be small compared with the magnetic one,
 $\e$ is the energy of the particle, $J_\perp=p_\perp^2/B$. Electromagnetic
fields are equal to the sum of stationary fields (without subscript) and
wave fields (subscript 1).
For cylindrical jet the stationary configuration of fields is (Istomin \&
Pariev, 1994) (c=1):
\def\omr{\Omega^F r}
$$ {\bf B}=B_z{\bf e}_z+B_\phi{\bf e}_\phi;\quad B_\phi=\omr B_z;$$
$$ {\bf u}=K{\bf B}+\omr{\bf e}_\phi;\eqno (2) $$
$$ {\bf E}=-\omr[{\bf e}_\phi{\times\bf B}], $$
 where $r,\phi,z$ are cylindrical coordinates. In components
$$ {\bf B}=(0,\omr,1)B_z,$$
$$ {\bf E}=(-\omr,0,0)B_z.$$
Here $\Omega^F$ and $K$ are functions of $r$, $B_z$ does not depend on $r$.
Stationary electric field, which has an absolute value of $\omr B_z$, is
not small compared to the  magnetic field when $\omr>1$, therefore we must
consider our equations in the frame moving with plasma where $E\equiv 0$.
There are a lot of such reference frames because there is an arbitrary
parameter $K(r)$ in the expression for the velocity of plasma
which determines the radial profile of longitudinal velocity.
$K(r)B_z$ is the velocity along the magnetic field which is not related
to the rotation. We choose, however, the velocity ${\bf u}$ which minimizes the
kinetic energy of the plasma in the stationary reference frame.
$$ {\bf u}=\frac{(0,\omr,-(\omr)^2)}{1+(\omr)^2}.\eqno (3)$$
It seems that the plasma moves with that velocity in real jets.
Fields in the wave calculated in terms of Lagrangian displacement $\xi$
are equal (here we drop out the phase
coefficient $\mbox{exp}(-i\omega t+ikz+im\phi)$)
$$ B_{r1}=iB_zF\xi;$$
$$ B_{\phi 1}=-B_z(\omr\frac{d\xi}{dr}+\xi\frac d{dr}(\omr)+\frac k S D);$$
$$ B_{z1}=B_z(-\frac{d\xi}{dr}-\frac\xi r+\frac m{rS} D);\eqno (4)$$
$$ E_{r1}=B_z(\omr\frac{d\xi}{dr}+\xi\frac d{dr}(\omr)-\frac\omega S D);$$
$$ E_{\phi 1}=-iB_z(\omega-m\Omega^F)\xi;$$
$$ E_{z1}=iB_z\omr(\omega+k)\xi;$$
where
$$ D=r\frac{d\xi}{dr}\left(\Omega^F(\omega+k)-\frac m{r^2}\right)-
\xi\left(\Omega^F(\omega+k)+\frac m{r^2}\right),$$
$$ F=k+m\Omega^F,$$
$$ S=\omega^2-k^2-m^2/r^2\mbox{.} $$
For Lagrangian displacement $\xi$ and dimensionless pressure disturbance
$p_*=4\pi P_1/B_z^2$ we have a system of differential
equations obtained by Istomin \& Pariev, 1996:
\def\fracb#1#2{\frac{\displaystyle #1}{\displaystyle #2}}
$$ \left\{\begin{array}{rcl}
     A\fracb1r\fracb d{dr}(r\xi)&=&C_1\xi-C_2p_*;\\
        A\fracb{dp_*}{dr}&=&C_3\xi-C_1p_*.\\
          \end{array}  \right.\eqno (5)
$$
Here
$$ C_1=\frac2r\left(m\Omega^F-(\omr)^2(\omega+k)\right);$$
$$ C_2=-\frac{\omega^2-k^2-m^2/r^2}{\omega+k};$$
$$ C_3=-(\omega+k)(A^2-4{\Omega^F}^2);$$
$$ A=k-\omega+2m\Omega^F-(\omr)^2(\omega+k).$$
Let us expand equations (5) near the Alfv\'en surface, i.e. near the 
point $r_A$
where $A(r_A)=0$. Denoting $x=r-r_A$ we get
$$
\left\{
\begin{array}{l}
\fracb{d\xi_r}{dx}=\fracb{1}{A'x}(C_1\xi_r-C_2p_*);\\
\fracb{dp_*}{dx}=\fracb{1}{A'x}(C_3\xi_r-C_1p_*),
\end{array}
\right.\eqno (6)
$$
where $\left.A'=\fracb{dA}{dr}\right|_{r=r_A} $.
The general solution of these equations is
$$ \left(\xi\atop p_*\right)=\alpha_1\left(C_2\ln x\atop C_1\ln x-A'\right)+
\alpha_2\left( C_2 \atop C_1\right) \eqno (7) $$
where $\alpha_1$ and $\alpha_2$ are constants. They are not arbitrary,
rather they are fixed by the boundary conditions on
$\xi$ and $p_*$ obtained from solving general equation (5).
We see that $\xi$ and $p_*$ have a logarithmic singularity at the
Alfv\'en point.
Later we adhere to our expansion over $r-r_A$,
assuming $r=r_A$, $A=0$ whenever
the quantities under consideration have no singularities at $r=r_A$.
Let us calculate $d{\bf r}/dt$ and $d\e/dt$ in the first
order by fields of the wave.
Remembering that ${\bf E}={\bf 0}$ in our reference
frame and knowing that terms with the energy $\e$ contain small parameter
which is the ratio of the Larmour radius to the jet radius or to the
characteristic
length of variations of zero order fields we can
simplify drift equations as follows,
$$\left(\frac{{d\bf r}}{dt}\right)_0=v_\|{\bf e}_z,$$
$$\left(\frac{{d\bf r}}{dt}\right)_1=v_\|\frac{{\bf B}_1}{B}-
                             v_\|\frac{{\bf B(B\cdot B}_1)}{B^3}+
                             \frac{[{\bf E}_1\times{\bf B}]}{B^2} \eqno (8)$$
Here subscript 0 denotes the zero order term and 1 denotes the first
order term.  The force of inertia is also present in the expression of ${d\bf
r}/{dt}$ (see Sivukhin, 1965), but the term with inertia force contains the
mass, has the same order as terms with energy and
is omitted consequently.
For ${d\e}/{dt}$ we will also not take into account the inertia force
because it is smaller than gradient terms such as
$\e v_\|^2[{\bf B\times(B\nabla)B}]/ecB^4.$
When we substitute ${\bf B}_1$ and ${\bf E}_1$ into (8) we get
$$\left(\frac{{dr}}{dt}\right)_1=(v_\|-1)\frac{i\xi F}{\gamma}, \eqno(8*) $$
where $\gamma=\sqrt{1+(\omr)^2}$ is the Lorentz factor corresponding to the
velocity of the moving frame.
$\xi$ depends on $z,\phi,t$ by $\exp\{-i\omega t+ikz+im\phi\}$.
The phase of the wave $\Psi$ which is ''seen'' by the particle
can be calculated
using  Lorentz transformation formulae
$ \Psi=-\frac F\gamma t'+\frac F\gamma z'$.
Assuming $ dz/dt=v_\| $ we get $\Psi=(v_\|-1)\frac F\gamma t'$. As we see
the coefficient for the phase coincides with the coefficient in (8*).
It confirms our calculations because $\xi$ is the Lagrangian displacement.
The question of the trajectory of the particle will be considered later.

Now we proceed with calculating ${d\e}/{dt}$.

$$ \frac{d\e}{dt}=e{\bf E}_1\cdot\left(\frac{{d\bf r}}{dt}\right)_0+
\frac{\e v_\perp}{2B}\frac{\partial B_1}{\partial t}.\eqno (9) $$
After substitution of expressions for fields into (9) we get
$$ \frac{d\e}{dt}=\e v_\perp^2\left(-\frac{iF}\gamma\right)\frac\xi{rS}
(\omega+k)[F-\omega],\eqno (10) $$
or, combining with (8*)
$$ \frac{d\e}{dr}=\frac{\e v_\perp^2}{1-v_\|}\left(\frac{(\omega+k)
[F-\omega]}{rS}\right).\eqno (11) $$
As we see, $\e$ is proportional to the $r$ if we do not take into account
dependence of the right hand side on $\e$
Therefore it makes sense to consider motion of the particle ''smoothed'' over
many periods. Taking the $r$-component of (8*), we have
$$\frac{dx}{dt}=i\omega^*\xi(x)e^{i\omega^*t},\quad \omega^*=(v_\|-1)\frac 
F\gamma. \eqno (12)$$
Near the Alfv\'en resonance $\xi(x)$ has a logarithmic singularity
$$\frac{dx}{dt}=\omega^*[A\ln x\sin(\omega^*t)+B\cos(\omega^*t)]. \eqno(13)$$
 It is clear that $\omega^*A$ and $\omega^*B$ are dimensionless amplitudes
of the perturbation \label{ampl}(remember, that $c$=1).
 Let us average equation~(13) expanding it in small amplitude of oscillation
of a particle in the field of the wave.
We denote perturbations of the zero, first, second orders as $x_0$, $x_1$,
$x_2$.
 $$\frac{dx_1}{dt}=\omega^*[A\ln x_0\sin(\omega^*t)+B\cos(\omega^*t)] $$
 $$x_1=-A\ln x_0\cos(\omega^*t)+B\sin(\omega^*t) $$
 $$\frac{dx_2}{dt}=\omega^*[A\ln|x_0-A\ln x_0\cos(\omega^*t)+
B\sin(\omega^*t)|\sin(\omega^*t)+B\cos(\omega^*t)] $$
 $$\overline{\frac{dx_2}{dt}}=\omega^*AB\frac 1{2x_0}$$
 Now let us replace $x_0$ and $x_2$ by $x$. We obtain smoothed
equation of motion
 $$ \frac{d\bar x}{dt}=\omega^*\frac {AB}{2x},$$
 which has the solution
 $$ \bar x=\sqrt{\omega^*ABt}. \eqno(14)$$

 From equations (11), (14) it follows that the particle gains energy drifting
along the $r$ axis. However , its acceleration rate is proportional
to the transverse momentum which remains constant in accordance with the drift
equations~(1)
$$ \frac{dp_\perp}{dt}\propto\frac{dJ_\perp}{dt}=0. $$
Here we do not take into account the change 
of the magnetic field connected with
the wave field because of $B_1\ll B$.
 However, because of  synchrotron losses $J_\perp$ and $p_\perp$ are 
monotonically
decreased according to  ($\displaystyle\omega_B=\omega_{B0}\frac{m_0c^2}{\e}$)
$$ \frac{dJ_\perp}{dt}=-\frac 43\omega_B^2\frac {r_e}c
\left(\frac\e{m_0c^2}\right)^3
   \left(1-\frac{v_\|^2}{c^2}\right)J_\perp, \eqno (15)$$
this  leads to the situation when $p_\|>p_\perp$.
Such anisotropic distribution can become unstable to the excitation of
electromagnetic waves which  results in isotropization of distribution function
in the momentum space.
 We will treat instability of plasma according to Mikhaylovskii, 1968.
Mikhaylovskii, 1968 showed that in the case $p_\|>p_\perp$
only perturbations with
$\kappa_z=0$ must be taken into account. The frequencies
of these perturbations with
\def\oureps{\varepsilon_{33}}
${\bf \kappa}\perp{\bf B}$ are the solutions of dispersion 
relation $\oureps-N^2=0$,
where $\oureps$ is the component of dielectric tensor along the magnetic field,
$N=c\kappa/\omega$. The contribution of $i$ sort of plasma
into $\oureps$ is given
by the general formula
\def\pa{\partial}
\def\om{\omega}
\def\omb{{\omega_B}}
 $$\oureps^{(i)}=\frac{4\pi e^2}\om\left<\sum_{n=-\infty}^\infty
  \frac{v_z J_n^2\left(\kappa v_\perp/\omb\right)}{\om-n\omb}
  \left[\frac{\pa F}{\pa p_z}-\frac{n\omb}\om\left(\frac{\pa F}{\pa p_z}-
  \frac{v_z}{v_\perp}\frac{\pa F}{\pa p_\perp}\right)\right]\right> 
\eqno (16) $$
 Here $\bigl<\dots\bigr>$ means
$\int\dots p_\perp dp_\perp dp_z$
and the distribution function $F$ is chosen so that
$\int F p_\perp dp_\perp dp_z=n_0$, $n_0$ is the equilibrium particle density.
From formula (16) it is seen that electron and positrons make an equal
contribution to the $\oureps$ because the sum is symmetric to the simultaneous
replacement of $n$ with $-n$ and of $e$ with $-e$.
After some calculations we get
  $$\oureps-1=\frac{4\pi e^2}{\om^2}\left<v_z\left[\frac{\pa F}{\pa p_z}-
  \left(1-J_0^2\left(\frac{\kappa v_\perp}{\omb}\right)\right)
  \frac{v_z}{v_\perp}\frac{\pa F}{\pa p_\perp}\right]\right>+
  2\om^2\left<\sum_{n=1}^\infty\frac{\frac{v_z^2}{v_\perp}
  \frac{\pa F}{\pa p_\perp}}{\om^2-n^2\om_B^2}
  J_n^2\left(\frac{\kappa v_\perp}{\omb}\right)\right>. \eqno (17) $$
Dispersion relation of the small oscillations has the form
  $$\frac{c^2\kappa^2}{4\pi e^2}=\frac{\om^2}{4\pi e^2}+
  \left<v_z\left[\frac{\pa F}{\pa p_z}-\left(1-J_0^2\right)
  \frac{v_z}{v_\perp}\frac{\pa F}{\pa p_\perp}\right]+
  2\om^2\sum_{n=1}^\infty\frac{\frac{v_z^2}{v_\perp}
  \frac{\pa F}{\pa p_\perp}}{\om^2-n^2\om_B^2}J_n^2\right>.\eqno (18)$$
Study of~(18) shows~(Mikhaylovskii, 1968) that instability arises at
small values of $\om$.
Therefore let us assume $\om$ to be small and find it
 $$\om^2=\frac
  {\frac{c^2\kappa^2}{4\pi e^2}-\left<v_z\frac{\pa F}{\pa p_z}\right>+\left<
   \left(1-J_0^2\right)\frac{v_z^2}{v_\perp}\frac{\pa F}{\pa p_\perp}\right>}
  {\frac 1{4\pi e^2}-2\sum\limits_{n=1}^\infty\left<\frac{v_z^2}{v_\perp}
  \frac{\pa F}{\pa p_\perp}J_n^2/n^2\om_B^2\right>}. \eqno (19)$$
For further calculations we assume that 
$\kappa v_\perp$ is of the order of $\omb$
and estimate expression~(19) taking into account only
first terms of expansion of Bessel function over its arguments.
 $$\om^2=\frac{c^2\kappa^2-4\pi e^2\left<
 v_z\frac{\pa F}{\pa p_z}-
 \frac12\left(\frac \kappa\omb\right)^2v_z^2v_\perp\frac{\pa F}{\pa p_\perp}
                              \right>}
 {1-4\pi e^2\left<\frac1{2\om_B^2}\left(\frac \kappa\omb\right)^2
 v_z^2v_\perp\frac{\pa F}{\pa p_\perp}\right>}. \eqno (20)$$
Let us take integrals in~(20) in parts remembering that $\omb\propto1/\gamma$.
Here $\gamma=\e/m_0c^2$. As was shown in~(Mikhaylovskii, 1968)
the maximum increment
is achieved at the values of $\kappa^2\simeq\left<\om_B^2/v_\perp^2\right>$.
We will see later that
instability at this $\kappa$ satisfies our requirements
$$ \om^2=\om_{B0}^2\frac{
   m_0^2c^2\frac{\om_{B0}^2}{\om_p^2}+
   \left<p_\perp^2\right>\left<\frac1{\gamma^3}\right>+
   \left<p_\perp^2\right>\left<\frac{p_\perp^2}{\gamma^3m_0^2c^2}\right>-
   \frac12\left(\left<\frac{p_z^2}{\gamma^3}\right>+
   \left<\frac{p_z^4}{\gamma^3m_0^2c^4}\right>+
   \left<\frac{p_z^2}{\gamma}\right>\right)}
  {\frac{\om_{B0}^2}{\om_p^2}\left<p_\perp^2\right>+
   \left<p_z^2\gamma\right>+
\frac12\left<\frac{p_z^2p_\perp^2}{\gamma m_0^2c^2}\right>}.\eqno(21)$$
 Here $\omega_p^2=4\pi ne^2/m$ is the plasma frequency squared.
 We continue our consideration by simplifying our assumption about the
distribution
function. Let all particles have equal $p_\perp$ and $p_\|$.
  This is rather rough approximation but it allows us to understand where
  the instability threshold is.
  So, remove all signs $<>$ and for the sake of simplicity let $m_0=c=1$.
$$ \om^2\approx\frac{\om_{B0}^2}{\gamma^2}
   \frac{\gamma^4/\beta+p_\perp^2+p_\perp^4-(p_z^2+p_z^4+\gamma^2p_z^2)/2}
   {\gamma^2/\beta+\gamma^2p_z^2+p_z^2p_\perp^2/2}, \eqno (22) $$
 where $\beta=4\pi nmc^2\gamma/B^2$ is the ratio of the particle pressure
 to the magnetic field pressure. Although $\beta$ depends on energy let
 it to be constant to know where the threshold of the resonance
is in the
 case of fixed energy. If we consider ultrarelativistic particles $p^2\ll p^4$,
 we will have for the ratio of $p_z$ to $p_\perp$ at which the
imaginary part of
 $\om$ arises
$$ \frac{p_z^2}{p_\perp^2}=\frac{4-\beta+\sqrt{\beta(17\beta-8)}}{
4(\beta-1)}. \eqno (23)$$
 We denote the threshold of stability $p_z/p_\perp=\alpha,\quad \alpha>1$
 (Expression~(23) is not quite valid for very large $\beta>8$ but we do not
 consider large $\beta$ here). $\mbox{Im}\>\omega$ arises
 when $p_z>\alpha p_{\perp}$.
 As we see from~(22) instability is fast because the increment of it is
 proportional to the cyclotron frequency.
 In our case it is the least
 typical time and it is less then the wave period by twelve orders of
 magnitude.
 Because of this fact we will always assume the distribution function
 to be such that $\mbox{Im}\om=0$.
 More concrete, we will consider  $p_z/p_\perp=\alpha=\mbox{const}$.
 Assuming constant pitch angle for all particles
 may be confusing because as a result of the synchrotron emission
 of these particles we will obtain sharp maximum of the
emission in directions close to this
 angle. But one has to remember that the emission must be integrated
 in the different parts of the jet, and always there will be  particles
found having
 velocities pointed to a given point in the sky, since in different
 parts of the jet we transit into a different reference frames.
 Thus the final result may be less rough than the intermediate.
Now we understand that ${dJ_\perp}/{dt}\ne 0$  because of the
anisotropic plasma instability
and instead of the equation
$dJ_\perp=\mbox{constant}$ we use $p_z/p_\perp=\alpha=\mbox{constant}$.
For the derivative of energy we write equation~(11) adding the synchrotron
losses.
$$ \frac{d\e}{dt}=\frac{\e v_\perp^2}{1-v_\|}\frac{(\omega+k)[F-\omega]}{rS}
v_x-\frac23\frac{p_\perp^2}{m_0}\om_{B0}^2r_e .\eqno (24) $$
Substituting $v_x$ from (14) we obtain
$$ \frac{d\e}{dt}=\frac\e{1+\alpha^2-\alpha\sqrt{1+\alpha^2}}
\frac{(\omega+k)[F-\omega]}{rS}\left(\frac{\om^*AB}{4t}\right)^{1/2}
-\frac23\frac{\e^2}{1+\alpha^2}\om_{B0}^2\frac{r_e}{m_0} .\eqno (25) $$
 Let us introduce some conventional signs
$$ Q=\frac1{1+\alpha^2-\alpha\sqrt{1+\alpha^2}}
     \frac{(\omega+k)[F-\omega]}{S};  $$
$$ \e_1=\frac34\frac{m_0}{1+2\alpha^2-2\alpha\sqrt{1+\alpha^2}}
\left(\frac{(\omega+k)[F-\omega]}S\right)^2
\frac{{\om^*}^2AB}{\om_{B0}^2r_er^2\om^*}. \eqno (26) $$
 Here $(\omega+k)[F-\omega]/S $ is the fixed number for a given values of
 wavenumbers of the perturbation $m$ and $k$.
It is large near the fast magnetosonic resonance
 surface $S=0$. The coefficient of
 $1/(\om_{B0}^2r_er^2\om^*)$ is a large number . In the case of an AGN jet
 $B\approx10^{-2}$G (Begelman at al., 1984), $r\approx 1$pc it equals
 approximately to the $10^4$.
 The value of $ {\om^*}^2AB $ as we noted after formula~(13) is the
 dimensionless
 amplitude of perturbation squared.
 Now taking into account our notations the solution for $\e$ has the form
 $$ \e=\left(\frac1{\e_1}\left(Q\frac xr-1\right)+\left(\frac1{\e_1}+
 \frac1{\e_0}\right)\exp\left\{-Q\frac xr\right\} \right)^{-1}. \eqno (27) $$
 Here $\e_0=\e(x=0) $. Let us denote $Qx/r=x'$. We assume that $x'$ can
 be large because of the quantity $Q$ is large.
 It is seen that first $\e$ exponentially increases and then
 decreases as $1/x'$.
 We consider an asymptotic behaviour of $\e$ in the case of different
 $x'$ and $\e_0$.
 In the case of $\e_0<\e_1$
 $$ \e=\left\{\begin{array}{rrr} \label{asym}
               \e_1/x',&\quad&x'e^{x'}>\e_1/\e_0\\
               \e_0e^{x'},&\quad&x'e^{x'}<\e_1/\e_0\\
                \end{array}\right.\eqno (27*) $$
 In the case of $\e_0>\e_1$
 $$ \e=\left\{\begin{array}{rrr}
               \e_1/x',&\quad&x'>1\\
               2\e_1/{x'}^2,&\quad&x'<1\\
                \end{array}\right. $$
We reject the case when $\e_0>\e_1$ assuming that the typical initial energy
is much less then $\e_1$.

\section{ Formation of the spectrum of accelerated particles}

Our goal is to calculate the distribution function
averaged over $x'$  knowing the distribution function at $x=0$.
We will consider the stationary case, when $\pa f/\pa t=0$.
Let the distribution function to be given at the point $x'_0$ as a function
of $\e'=\e/\e_1$. Then the number of particles
$$dN=f(\e'_0,x'_0)d\e'_0dx'_0=f\Bigl(\e'_0(\e',x'),x'_0(\e',x')\Bigr)
 \frac{D(\e'_0,x'_0)}{D(\e',x')}d\e'dx'. $$
Further we omit primes.
Thus if we know the trajectory $\e_0(\e,x),x_0(\e,x)$ we may
obtain the distribution function at the point $\e,x$.
In our case the Jacobian $D(\e'_0,x'_0)/D(\e',x')$ is not equal to unity
because variables $\e,x$ are not canonical.
 So,
$$ F(\e,x)=f\Bigl(\e _0(\e ,x ),x _0(\e ,x )\Bigr)
 \frac{D(\e _0,x _0)}{D(\e ,x )}. \eqno (29) $$
 The formula  analogous to the (27) but with the arbitrary reference point
 $x_0$ is the following
 $$\e=\{x-1+(1/\e(x_0)-x_0+1)\exp(-x+x_0)\}^{-1} \eqno(30) $$
Substituting $\e_0$ expressed from (30) into (29) we obtain
\def\der#1#2{\frac{{\pa #1}}{{\pa #2}}}
$$ F(\e,x)=f\left(\left\{-1+x_0+\left(\frac1\e-x+1\right)\exp(x-x_0)
\right\}^{-1},x_0\right)
   \left(\der{\e_0}\e\der{x_0}x-\der{\e_0}x\der{x_0}\e\right)= \eqno(31)$$
$$ f\left(\left\{-1+x_0+\left(\frac1\e-x+1\right)\exp(x-x_0)\right\}^{-1},
x_0\right)
   \frac{\e_0^2}{\e^2}\exp(x-x_0)\frac x{x_0} $$
We know that initial fast particles are born near the point $x=0$ 
\label{init}, therefore
in the last formula we let $x_0=0$ everywhere except $x_0$ in the denominator.
Physically $f(\e_0,0)$ is the particle density at zero point and $x_0$ is
the distance at which our formulae, for instance~(14), do not work.
Because of the small imaginary part of $\omega$
which is an attenuation increment of the wave~(Istomin \& Pariev, 1996)
the divergence $\ln x$ at the resonance is cut at the point
$$ r_0\approx\left|\frac{A(r=r_A)}{dA/dr(r=r_A)}\right|.\eqno (31*)$$
Here $A(r=r_A)$ is not equal to zero as we assumed in the previous text,
but determined by the small value of the imaginary part of $\omega$.
Thus, $x_0$ is the width of the resonance $ x_0=Qr_0/r_A$.
Eventually,
$$ F(\e,x)= f_0\left(\left\{-1+\left(\frac1\e-x+1\right)\exp(x)\right\}^{-1}
\right)
   \frac{\e_0^2}{\e^2}\exp(x)\frac x{x_0}  \eqno (32) $$
Because the width of the resonance $r_0$ and the width of the
acceleration region $\Delta r$ \label{width}, which is  some function of $r_0$,
are small compared
with the jet radius we average the distribution function over the  jet radius
near the resonance surface
$$ \bar f(\e)=\frac1{\Delta x}\int\limits_0^\infty F(\e,x)\,dx,\quad 
\Delta x=\frac Q{r_A}\Delta r. \eqno (34) $$
Denoting $N(\e)=\bar f(\e)\Delta x$ we obtain

$$ \bar N(\e)=\int\limits_0^\infty f_0\left(\left\{-1+\left(\frac1\e-x+
1\right)\exp(x)\right\}^{-1}\right)
   \frac{\e_0^2}{\e^2}\exp(x)\frac x{x_0}\,dx. \eqno(35) $$
Let us change variable of integration to $\e_0$.
The dependence of $\e_0$ versus $x$ under fixed $\e$ is presented in Fig.~1.
\begin{figure}
\vspace{8cm}
\caption{The dependence of $\e_0$ versus $x$
(the energy $\e$ and the distance $x$
are in the values of $\e_1$ and $r/Q$ (26) correspondently)  }
\label{fig1}
\end{figure}

The value of $\e_{min}$ is
$$ \e_{min}=1/(-1+exp(1/\e)). \eqno (36*)$$
Because the function $x=x(\e_0)$ has two branches the equation~(35)
is transformed into
\def\e{{\cal E}}
$$ N(\e)=\frac1{\e^2x_0}\int\limits_{\e_{min}}^\e f_0(\e_0)\frac x{1/\e-x}
\,d\e_0+$$
$$ \frac1{\e^2x_0}\int\limits_{\e_{min}}^\infty f_0(\e_0)\frac x{1/\e-x}
\,d\e_0,\eqno(36)$$
 where in the first integral $x$ is taken on the first branch of $x(\e_0)$
and in the second integral on the second branch.
 Let us make some natural assumptions.
 Initial distribution function is cut off at the typical energy of ${T_0}$,
provided that $T_0 \ll 1$ (bearing in mind that $T_0$ as well as $\e$ are
in the
units of energy $\e_1$~(26)).
Let us calculate $N(\e)$ in the region, where $T_0\ll\e\ll 1$, because
it is seen
from equations~(36) and (36*) that in the case $\e\sim1$ the integral is small.
 Let us evaluate the contribution of the singularity $x=1/\e$ in integrals~(36)
$$ \e_0-\e_{min}=\frac{d^2\e_0}{dx^2}\left(x-\frac1{\e}\right)^2 $$
Integral near the singularity is
$$ \frac1{\e^2} \int\limits_{\e_{min}} f_0(\e_{min})\frac{\frac1{\e}\sqrt{
   \frac{d^2\e_0}{dx^2}}}{x_0\sqrt{\e_0-\e_{min}}}\,d\e_0\sim\sqrt{\e_0-
\e_{min}}. $$
Thus, its contribution is small. Because the main contribution to
expression~(30) is when $x$ is far
from $1/\e$, we may use asymptotic values of $x$ (see equation~(27*)):
$ x\approx\ln(\e/\e_0) $  on the first branch;
$ x\approx1/\e+1 $ on the second branch and obtain that
$$ N(\e)=\frac1{\e^2x_0}\int\limits_{\e_{min}}^\e f_0(\e_0)
\frac{\ln(\e/\e_0)}{1/\e-\ln(\e_0/\e)}\,d\e_0
+\frac1{\e^2x_0}\int\limits_{\e_{min}}^\infty f_0(\e_0)\frac 1\e
\,d\e_0.\eqno(37)$$
Let us consider only $\e_{min}\ll T_0$ i.e. $\e\ll-1/\ln T_0$
because in the reverse case
the integral is small. Now we can replace $\e_{min}$ by~$0$
and also $1/\e\gg\ln(\e/\e_0)$. Equation~(37) becomes
$$ N(\e)=\frac1{\e x_0}\int\limits_0^\e f_0(\e_0)(-\ln \e_0)\,d\e_0+
\frac n{\e^3 x_0}, \eqno(38) $$
where $n=\int_0^\infty f_0(\e_0)\,d\e_0$. Rewriting expression~(38) as
$$ N(\e)=\frac n{\e x_0}<-\ln \e_0>+ \frac n{\e^3 x_0} \eqno (39) $$
we note that the first term is approximately equal to
$$ \frac n{\e x_0} (-\ln T_0) $$
and is small compared to the second term because
non-equality $1/\e^2 \gg -\ln T_0$ is the
consequence of non-equality $1/\e \gg -\ln T_0$.
Therefore $N(\e)\propto 1/\e^3 $ and we obtain power law distribution with
the index~$-3$. The validity region of this formula $T_0<\e<-1/\ln T_0$ 
is quite
large.

We performed integration in equation~(36) numerically having taken Gaussian
initial distribution with
different temperatures. Results are presented in Figs.~2 and~3.

\begin{figure}
\vspace{8cm}
\caption{Distribution function calculated with Gaussian initial
distribution with temperature $2\cdot10^{-5}\e_1$ (the energy $\e$
is in the values of $\e_1$ (26))
}
\label{fig2}
\end{figure}

\begin{figure}
\vspace{8cm}
\caption{Distribution function calculated with Gaussian initial distribution
with temperature $10^{-3}$ (the energy $\e$
is in the values of $\e_1$ (26))
}
\label{fig3}
\end{figure}

The value of power law
index is minus three for small energies and rises slightly before
the cut off of the distribution function.
However, the value $-3$ of power law index is not applicable all the way down
to $\e=0$.
The natural boundary is the width of the acceleration region $\Delta r$
(see expression~(31*)). Because the main contribution to the integral~(31*)
is from the region near the
point $x=1/\e+1$ the lower limit for power law $1/\e^3$ is $\e_{l}=1/\Delta x$.
We also found mean power law index
for the distribution of particles between energies $\e_l$
and $-1/\ln T_0$. Results are presented in Fig.~4.

\begin{figure}
\vspace{8cm}
\caption{Averaged power law index (the initial particles energy $T_0$
is in the values of $\e_1$ (26))
}
\label{fig4}
\end{figure}

One remaining question in the formation of the spectrum of
energetic particles is the origin
\label{origin}of the initial particles, since in the first
order on perturbed fields there is no acceleration of
the cold plasma in equations~(1).
We have already mentioned that initial particles to be accelerated
are originated
in the region close to the Alfv\'en resonance.
It can be seen from equations~(1) that
in the second order on perturbed fields
the acceleration always takes place and the power
acquired corresponds to the term ${\bf j\cdot E}$ in the energy conservation
equation for the electromagnetic field. The damping of the wave occured due
to the heating of the plasma by current ${\bf j}_1$ which is described
by second-order terms.
If all energy of the wave is transmitted to particles, the condition
$\beta>1$ would imply that the energy density of the wave is
more than the energy
density of the stationary field. Let this to be true at least on the surface
of the resonance
where the field is maximal. Then the energy acquired by one particle can be
evaluated as $T_0\approx B_0^2/8 \pi n$.
Given typical values $B_0\approx 10^{-2} \mbox{G}$ 
and $n\approx 0.1 \mbox{cm}^{-3}$ we
obtain $T_0\approx 10\mbox{MeV}$.

\section{ Synchrotron emission. }

Let us now consider synchrotron emission of accelerated particles in a
cylindrical jet with spiral magnetic field.
Consistently with the results of section~3
we take isotropic distribution function of emitting particle in momentum
and power law in energy
\def\g{\gamma}
$$ dn=K_e\e^{-\g}d\e dVd\Omega_{\bf p} \eqno (40) $$
Here $dn$ is the number of particles in the energy interval
$\e,\e+d\e$, $dV$ is elementary volume,
$d\Omega_{\bf p}$ is the elementary solid angle in the direction of the
particle momentum ${\bf p}$, $K_e=\mbox{constant}$, $\gamma=\mbox{constant}$.
We will assume that ultralelativistic particles are distributed according (40)
in the frame moving with plasma with the velocity
$$ {\bf u}(r)=K{\bf B}+\Omega^F(r)r{\bf e}_{\phi} $$
We will consider two cases for the spatial distribution of particles, namely,
uniform in the jet volume and concentrated close to the Alfv\'en resonant
surface $r=r_A$.
Influence of plasma on the emission of relativistic $e^-$ and $e^+$ is
neglected. Absorption and reabsorption are also not taken into account.
 At first we assume that there is no Doppler boosting i.e.
 ${\bf u}=0$. In this case, according to e.g. Ginzburg, 1989, Stocks parameters
are
\def\chit{\tilde\chi}
$$ I=\frac{\g+7/3}{\g+1}k(\nu)\int(B\sin\chi)^{1/2(\g+1)}\,dR,$$
$$ Q=k(\nu)\int(B\sin\chi)^{1/2(\g+1)}\cos 2\chit\,dR,                
\eqno(41)$$
$$ U=k(\nu)\int(B\sin\chi)^{1/2(\g+1)}\sin 2\chit\,dR,$$
$$ V=0, $$
Here integration is to be fulfilled along the line of sight in the region
filled with particles,
$$ k(\nu)=\frac{\sqrt3}4\Gamma\left(\frac{3\g-1}{12}\right)
 \Gamma\left(\frac{3\g+7}{12}\right)\frac{e^3}{mc^2}\left[
   \frac{3e}{2\pi m^3c^5}\right]^{1/2(\g-1)}\nu^{-1/2(\g-1)}\frac{K_e}{D^2},$$
 $\chi$ is the angle between the magnetic field ${\bf B}$ and the line of
 sight, $\chit$ is the position angle in the picture plane of the electric
 field vector, $D$ is the distance between the source and the observer.
We will measure $\chit$ clockwise from the direction parallel
 to the jet axis.
For ultrarelativistic particles $V=0$, and the emission has only linear
polarization. Degree of polarization of the outgoing radiation is expressed
as $\Pi=\sqrt{Q^2+U^2}/I $, resultant position angle of the electric field is
found from
$$ \cos 2\chit_{res}=\frac Q{\sqrt{Q^2+U^2}},
\quad \sin 2\chit_{res}=\frac U{\sqrt{Q^2+U^2}},
\quad 0\le\chit_{res}<\pi.$$
Let $h$ be the distance from the line of sight to the jet axis.
We will assume that
$h$ is signed, i.e changing from $-R$ to $R$ for the different lines of sight.
We will denote the angle between the jet axis and the direction
to the observer as~$\theta$.
In cylindrical coordinates
$$ I=\frac{\g+7/3}{\g+1}k(\nu)\int\limits_{\phi_1}^{\phi_2} |B_\perp|^{1/2(\g
+1)}\frac{h}{\sin\theta\sin^2\phi}\,d\phi,\eqno(42) $$
$$ Q=k(\nu)\int\limits_{\phi_1}^{\phi_2}\frac{|B_\perp|^{1/2(\g+1)}}{|
B_\perp|^2}\frac{[B_\phi^2\cos^2\phi-
(B_z\sin\theta+B_\phi\sin\phi\cos\theta)^2]h}{\sin\theta\sin^2\phi}\,d\phi,
\eqno(43) $$
$$ U=k(\nu)\int\limits_{\phi_1}^{\phi_2}\frac{|B_\perp|^{1/2(\g+1)}}{
|B_\perp|^2}2B_\phi\cos\phi
(B_z\sin\theta+B_\phi\sin\phi\cos\theta)\frac{-h}{\sin\theta\sin^2\phi}
\,d\phi,$$
where $\displaystyle \phi_1=\pi\mbox{sgn}\,h-\arcsin\frac{h}{R},
\quad\phi_2=\arcsin \frac{h}{R} $.
It is seen that when one changes $\phi$ by $\pi-\phi$ the expression for $U$
changes its sign and, therefore, $U=0$.
Consequently, if $Q>0$ than $\chit_{res}=0$,
if $Q<0$ then $\chit_{res}=\pi/2$. Thus, the electric field vector
either parallel
or perpendicular to the jet axis. This fact may account for the
bimodal distribution of observed $\chit_{res}$ (Gabuzda et al., 1994a).
In the case
$B_z=\mbox{constant},\quad B_\phi=B_z\Omega^F r$ and $\g=3$ expressions for
$I$ and $U$ from the unit jet length (i.e. (42) and (43) integrated over~$h$)
$$ I=\frac{\g+7/3}{\g+1}k(\nu)\int\limits_{-R}^R dh\int\limits_{\phi_1}^{
\phi_2} |B_\perp|^{1/2(\g+1)}\frac{h}{\sin\theta\sin^2\phi}\,d\phi,\eqno(44) $$
$$ Q=k(\nu)\int\limits_{-R}^R dh\int\limits_{\phi_1}^{\phi_2}|B_\perp|^{1/2
(\g-3)}\frac{[{\Omega^F}^2r^2\cos^2\phi-
(\sin\theta+\Omega^F r\sin\phi\cos\theta)^2]B_z^2h}{\sin\theta\sin^2\phi}
\,d\phi,\eqno(45) $$
may be reduced to the form
$$ I=\frac{\g+7/3}{\g+1}k(\nu)B_z^{(\g+1)/2}
\frac\pi{\sin\theta}\int\limits_0^R2t\left[\sin^2\theta+{\Omega^F}^2r^2(t)(1-
\frac12\sin^2\theta)\right]\,dt, $$
$$ Q=k(\nu)B_z^{(\g+1)/2}\pi\sin\theta\int\limits_0^R({\Omega^F}^2r^2(t)-2)t
\,dt. $$
From above we see that in the case $\g=3$ the sign of $Q$ does not depend
on $\theta$, but determined only by function $\Omega^F(t)$.
For the degree of polarization
$$ \Pi=\frac{\g+1}{\g+7/3}\sin^2\theta\frac{\int\limits_0^R({\Omega^F}
^2r^2(t)-2)t\,dt}
   {\int\limits_0^R 2t\,dt(\sin^2\theta+{\Omega^F}^2r^2(t)(1-\frac12\sin^2
\theta))} .$$
In the case $\g=3$ and $\Omega^F=\Omega(1-(t/R)^2)$
$$ \Pi=\frac{\displaystyle\frac{1}{2}\left(\frac{\Omega^2}{12}-2\right)}
{\displaystyle
\sin^2\theta+\frac{\Omega^2}{12}\left(1-\frac{1}{2}\sin^2\theta\right)}
     \frac34\sin^2\theta $$
If $|\Omega|>2\sqrt6$, then $Q>0$, $\chit_{res}=0$, i.e. orientation of the 
magnetic field in the jet derived from observations is
perpendicular to the jet axis. If $0<|\Omega|<2\sqrt6$ magnetic field
looks parallel to the jet axis. Equations (44) and (45) were integrated
numerically for
$ \Omega^F=\Omega(1-(t/R)^2)$ and $\g=2.1$, which correspond to the power law
index in the spectrum of synchrotron emission of $(\g-1)/2=0.55$, the mean
observed value for extragalactic jets.
Results are presented in Fig.~5. Positive values of the degree of
polarization
correspond to $\chit_{res}=0$, negative to  $\chit_{res}=\pi/2$. In the case
$\Pi=0$ radiation is entirely unpolarized. Note also that in nonrelativistic
consideration $\Pi(180^0-\theta)=\Pi(\theta)$ for any value of $\theta$.

\begin{figure}
\vspace{10cm}
\caption{Dependences of linear polarization $\Pi$ on angular
rotational velocity $\Omega$~(a) and on visual velocity of
knots~$v_{vis}$~(b) for fundamental axisymmetric mode $m=0$.
$\Pi>0$ when electric field
vector is oriented parallel
to the jet axis, $\Pi<0$ when electric field vector is oriented
perpendicular to the jet axis.
$v_{vis}>0$ means that observed motion of
knots is direct, $v_{vis}<0$ means reversal motion.
Crosses are for BL~Lac objects and
squares are for quasars listed in Table~1.}
\label{fig5}
\end{figure}

In paper Istomin~\& Pariev, 1996 we presented a hypothesis that knots observed
in jets can be the manifestation of the  ''standing wave'' phenomenon.
If one knows the dispersion curves $\om=\om(k)$ and wavenumbers~$k_{min}$ at
which $d\om/dk=0$
for perturbations one can calculate the velocity of moving crests of the
''standing wave'' as
$ v_{phas}=\om_{min}(k_{min})/k_{min}$
and corresponding observed velocity $\displaystyle v_{vis}=\frac{v_{phas}
\sin\theta}
{1-v_{phas}/c\cos\theta}$.
Note also that changing sign of~$B_{\phi}$ in~(2) is equivalent to
changing sign of $\Omega^F$, since $I$ and $Q$ depend only on $B_z$ and
$B_{\phi}$ and not on $u$ or $E_r$.
 On Fig.~5 we present the relationships
of the polarization $\Pi$ of jet emission and $v_{vis}$ obtained in the case
of $\Omega$ changing from $0$ to $100$ for some angles $\theta$.
We also plotted observational points for jets in
BL Lac objcts and quasars for which
we were able to find both polarization and proper motion measurements in
the literature and which are listed in Table~1.
For each object we calculated degree
of polarization and observed velocity as averaged over all bright knots
observed in a particular object excluding a few knots peculiarity of which
was indicated in original references. This pocedure might be very questionable
although in the lack of observations of the emission from the space between
knots on parsec scale we had nothing better. Numbers for $v_{vis}$ were
calculated assuming Hubble constant $H=100h\mbox{km/s}\cdot\mbox{Mpc}$
\begin{table}
\caption{Degree of polarization and velocities of knots in BL~Lac objects
and quasars}

\begin{tabular}{llccc}
Class & Source & $\Pi\,$,\% & $v_{vis}h$ & References \\
\hline
BL~Lac & 0454+844 & 25 & 0.6 & 2,3\\
BL~Lac & 0735+178 & 14.2 & 5.5 & 4,5\\
BL~Lac & 0954+658 & 17.5 & 5.7 & 1\\
BL~Lac & 1803+784 & 11.5 & 1.8 & 1\\
BL~Lac & 1823+568 & 17.75& 4.05& 1\\
BL~Lac & 2007+777 & 5.55 & 2.9 & 2,1\\
quasar,& 1308+326 & 3.77 & 10.9 & 6\\
BL~Lac?&&&&  \\
BL~Lac & 0851+202 & 6.2 & 2.8 & 7\\
BL~Lac & 1219+285 & 30 & 2.0 & 7\\
BL~Lac & 2200+420 & 10.9 & 3.5 & 7\\
quasar & 0212+735 &-13.35 & 3.9 & 8\\
quasar & 0836+710 &-31.33 & 7.8 & 8\\
quasar & 3C279    &-15.37 & 5.6 & 8\\
quasar & 3C345    &-7. & 7.7 & 9\\
quasar & 1928+738 &-9.15 & 5.78 & 8\\
quasar & 3C454.3  &-15.55 & 8.9 & 8
\end{tabular}

\medskip
References.--- (1)~Gabuzda et al. (1994a), (2)~Witzel et al. (1988),
(3)~Gabuzda et al. (1989), (4)~B\aa\aa~th \& Zhang (1991),
(5)~Gabuzda et al. (1994b), (6)~Gabuzda \& Cawthorne (1993),
(7)~Gabuzda \& Cawthorne (1996), (8)~private communication of Gabuzda~D.C.,
(9)~Brown, Roberts, \& Wardle (1994)
\end{table}

In the case of accelerated particles concentrated very close to the Alfv\'en
surface $r=r_A$ one can obtain from the expressions~(44) and (45) the
following formulae
\begin{eqnarray}
I & = & \frac{\gamma+7/3}{\gamma+1}k(\nu)B_z^{\frac{\gamma+1}{2}}\int
\limits_0^{2\pi} \frac{r_A}{\sin\theta}\left[\sin^2\theta+
{\Omega^F}^2(r_A)r_A^2(1-\sin^2\phi\sin^2\theta)+2\Omega^F r_A(t)\sin\phi
\cos\theta\sin\theta\right]^{\frac{\gamma+1}{4}}\,d\phi\mbox{,} \nonumber\\
Q & = & k(\nu)B_z^{\frac{\gamma+1}{2}}
\int\limits_0^{2\pi}\frac{r_A}{\sin\theta}
\left[\sin^2\theta+{\Omega^F}^2(r_A)r_A^2(1-\sin^2\phi\sin^2\theta)
+2\Omega^F(r_A)r_A\sin\phi\cos\theta\sin\theta\right]^{\frac{\gamma-3}{4}}
\times   \nonumber\\
&& \left({\Omega^F}^2(r_A)r_A^2\cos^2\phi-(\sin\theta+\Omega^F(r_A)r_A\sin\phi
\cos\theta)^2\right)\,d\phi\mbox{.}  \nonumber
\end{eqnarray}

\begin{figure}
\vspace{8cm}
\caption{The same plots as in Fig.~5 but emitting particles are concentrated
close to the Alfv\'en resonance surface, $m=1$ fundamental mode.}
\label{fig6}
\end{figure}

Since for axisymmetric modes ($m=0$) always
$|\omega_{min}|>|k_{min}|$~(Istomin~\& Pariev, 1994) one can see
from the expression~(5)
for $A$ that Alfv\'en resonance does not exist for axisymmetric modes.
Therefore, there should be no particle acceleration by this mode. The next
most important modes are $m=1$ and $m=-1$. Since no one quantity $k_{min}$,
$\omega_{min}$ or $\Pi$ changes when~$m$ changes to~$-m$ and $\Omega^F$ to
$-\Omega^F$, it is enough to consider only $m=1$ and all values of $\Omega^F$.
Polarization $\Pi$ versus $v_{phas}$ for this case is plotted
in Fig.~6
as well as observational points. We showed first non-axisymmetric
fundamental mode $m=1$ which are expected to be excited more than
higher modes with $m>1$. Polarization curves are plotted only for values
of $\Omega$ at which Alfv\'en resonance $r=r_A$ is present inside the
jet. As previously, $\Omega^F=\Omega(1-(r/R)^2)$ and $\gamma=2.1$.
It is seen that in the case of uniform particle distribution across the
jet, curves for $\theta$ changing from $120^\circ$ to $70^\circ$  are
located in the region where points from observations of
BL~Lac object (crosses) and quasars (squares) are situated. However,
polarization curves for emitting particles concentrated on the Alfv\'en
resonance surface, as it is the case if our acceleration mechanism works,
do not match observation of BL~Lac objects. Electric field vector is
always oriented perpendicular to the jet axis.

Now let us make calculations closer to the reality taking into account
Doppler boosting and presence of strong electric field.
We will denote values measured in the plasma rest frame as values with
primes. Let us find the transformation formulae for the polarization tensor
$J_{\alpha\beta}=<E_\alpha E_\beta^*>$ and Stocks parameters
$J_{\alpha\beta}=\frac12\left(\begin{array}{cc}
                              I+Q & U\\
                              U & I-Q
                              \end{array}\right)$
when expressed in the source frame $x_sy_sz_s$ and the observer frame
$x_oy_oz_o$. Using the transformation formulae for the electric and
magnetic fields of the wave directly it is possible to
obtain
$$ J_{\alpha\beta o}=J^\prime_{\alpha\beta s}
\frac{1-u^2/c^2}{(1-u/c\cos\delta)^2}. \eqno (46)$$
Here we denote by $\delta$ an angle between velocity ${\bf u}$ of the
source and the direction towards observer~${\bf k}$ measured in the observer
frame, correspondently $\delta'$ is the same angle but measured in the
source frame.
Thus, polarization ellipse in the observer frame is obtained
from polarization ellipse in the source frame by the rotation by the angle 
$\delta-\delta'$ in the plane containing the direction to the
observer ${\bf n}$
and ${\bf u}$. Therefore the degree of polarization is not changed.
Since for one particle
Stocks parameters are expressed through the radiation flux with two
main directions of polarization $p^{(1)}_{\nu'}$ and
$p^{(2)}_{\nu'}$ by the formulae $I=p^{(1)}_{\nu'}+p^{(2)}_{\nu'}$,
$Q=(p^{(1)}_{\nu'}-p^{(2)}_{\nu'})\cos2\chit$ ,
$U=(p^{(1)}_{\nu'}-p^{(2)}_{\nu'})\sin2\chit$ then $p^{(1)}_{\nu}$ and
$p^{(2)}_{\nu}$ are transformed as
$\displaystyle p^{(1)}_{\nu}=p^{(1)}_{\nu'}\frac{d\nu'}{d\nu}\frac{1-
u^2/c^2}{(1-u/c\cos\delta)^2}$,
$\displaystyle p^{(2)}_{\nu}=p^{(2)}_{\nu'}\frac{d\nu'}{d\nu}\frac{1-
u^2/c^2}{(1-u/c\cos\delta)^2}$.
Therefore the intensity
$$ I(\nu,{\bf k})=\int(p^{(1)}_{\nu'}(\nu',\e',{\bf R},\chi',\psi')+
                       p^{(1)}_{\nu'}(\nu',\e',{\bf R},\chi',\psi'))\times $$
$$
\frac{d\nu'}{d\nu}\frac{1-u^2/c^2}{(1-u/c\cos\delta)^2}K_e{\e'}^{-\g}
\frac{dV'}{dV}d\e'd\Omega_{\tau'}R^2dR, $$
where $\chi'$ is an angle between momentum of the particle and the
magnetic field vector, $\psi'$ is the angle between the direction of
radiation and the cone formed by rotating velocity vector, both measured
in the frame comoving with the fluid element.
We will make integration over momenta in the plasma rest frame while
integration over space coordinates in the observer frame.
Integration over solid angle $\Omega_{\tau'}$ in the momenta space
may be done analogous to the usual case of synchrotron emission of
ultra-relativistic particles,
when there is no electric field ${\bf E=-u\times B}$. Integration over
$\e'$ leads to (see, e.g., Ginzburg, 1989)
$$ I(\nu,{\bf k})=\frac{\gamma+7/3}{\gamma+1}k(\nu)
\int \left(\frac{1-u/c\cos\delta}{\sqrt{1-u^2/c^2}}\right)^{-\frac{\gamma-
1}{2}}(B'\sin\chi')^{\frac{\gamma+1}{2}}
\frac{dR}{1-u/c\cos\delta} \eqno (47) $$
Here $\displaystyle\nu'=\nu\frac{1-u/c\cos\delta}{\sqrt{1-u^2/c^2}}$,
$\delta$ is the
angle between ${\bf u}$ and the direction to the observer, $\chi'$ is the
angle between the magnetic field in the plasma rest frame ${\bf B'}$ and the
line of sight, integration is performed along the line of sight.
The expressions
for $Q$ and $U$ are obtained in the same fashion.
$$ Q(\nu,{\bf k})=k(\nu)
\int {\left(\frac{1-u/c\cos\delta}{\sqrt{1-u^2/c^2}}\right)}^{-\frac{\gamma-
1}{2}}(B'\sin\chi')^{\frac{\gamma+1}{2}}
\frac{\cos 2\chit\, dR}{1-u/c\cos\delta},$$
$$ U(\nu,{\bf k})=k(\nu)\int {\left(\frac{1-u/c\cos\delta}{\sqrt{1-u^2/c^2}}
\right)}^{-\frac{\gamma-1}{2}}(B'\sin\chi')^{\frac{\gamma+1}{2}}
\frac{\sin 2\chit\, dR}{1-u/c\cos\delta}.$$
Using the transformation  rules leads us to the expression for the
$B'\sin\chi'$
$$ B'^2\sin^2\chi'=B^2_\phi+B_z^2-E_r^2-$$
\nopagebreak
$$
\frac{[-E_r(1-u^2)(u_z\sin\phi\sin\theta+u_\phi\cos\theta)+u(\cos\delta-
u)(u_\phi B_\phi+u_zB_z)]^2}
{u^4(1-u\cos\delta)^2} \eqno (48) $$
and using of (46) gives us the expression for $\chit$
$$ \chit=\chit_u+\sigma,\eqno (49)$$
$$ \tan\chit_v=\frac{E_r(u^2-u\cos\delta)(u_z\sin\phi\sin\theta+u_\phi
\cos\theta)-
   u^2\sin^2\delta(u_\phi B_\phi+u_z B_z)}
  {E_ru^2\sin\theta\cos\phi(1-u\cos\delta)},$$
$$ 0\le\chit_u<\pi, $$
$$ \cos\sigma=-\frac{u_\phi\sin\phi\cos\theta+u_z\sin\theta}{u\sin\delta},
\quad \sin\sigma=-\frac{u_\phi\cos\phi}{u\sin\delta}, $$
$$ u\cos\delta=-\sin\theta\sin\phi u_\phi+\cos\theta u_z $$
Assuming uniform particle distribution for the emission from the unit
length we obtain
$$
I=\frac{\gamma+7/3}{\gamma+1}\frac{k(\nu)}{\sin\theta}\int_{0}^R r\,dr
\int_0^{2\pi}\,d\phi
\left(\frac{1-u\cos\delta}{\sqrt{1-u^2}}\right)^{-\frac{\g-1}{2}}
\frac{|B'\sin\chi'|^{(\g+1)/2}}{1-u\cos\delta}$$
$$
Q=\frac{k(\nu)}{\sin\theta}\int_{0}^R r\,dr\int_0^{2\pi}\,d\phi
\left(\frac{1-u\cos\delta}{\sqrt{1-u^2}}\right)^{-\frac{\g-1}{2}}
|B'\sin\chi'|^{(\g+1)/2}\frac{\cos2\chit}{1-u\cos\delta}\,\eqno (50)$$
$$
U=\frac{k(\nu)}{\sin\theta}\int_{0}^R r\,dr\int_0^{2\pi}\,d\phi
\left(\frac{1-u\cos\delta}{\sqrt{1-u^2}}\right)^{-\frac{\g-1}{2}}
|B'\sin\chi'|^{(\g+1)/2}\frac{\sin2\chit}{1-u\cos\delta}\mbox{.}$$
Assuming particles to be located close to the Alfv\'en resonance
surface in the region $|r-r_A|\le\Delta$ and $\Delta\ll R$  we have
$$
I=\frac{\gamma+7/3}{\gamma+1}\frac{k(\nu)}{\sin\theta}\,\Delta
\int_0^{2\pi}\,d\phi r_A
\left(\frac{1-u\cos\delta}{\sqrt{1-u^2}}\right)^{-\frac{\g-1}{2}}
\frac{|B'\sin\chi'|^{(\g+1)/2}}{1-u\cos\delta}$$
$$
Q=\frac{k(\nu)}{\sin\theta}\,\Delta\int_0^{2\pi}\,d\phi r_A
\left(\frac{1-u\cos\delta}{\sqrt{1-u^2}}\right)^{-\frac{\g-1}{2}}\times$$
$$
|B'\sin\chi'|^{(\g+1)/2}\frac{\cos2\chit}{1-u\cos\delta}\,\eqno (51)$$
$$
U=\frac{k(\nu)}{\sin\theta}\,\Delta\int_0^{2\pi}\,d\phi r_A
\left(\frac{1-u\cos\delta}{\sqrt{1-u^2}}\right)^{-\frac{\g-1}{2}}\times$$
$$
|B'\sin\chi'|^{(\g+1)/2}\frac{\sin2\chit}{1-u\cos\delta}\mbox{.}$$
In the force--free approach constant $K$ determining the component of ${\bf u}$
parallel to the magnetic field in the observer frame ${\bf B}$ remains
to be free. We choose it such that the velocity ${\bf u}$ of plasma
has the least possible
absolute value for a given $B_z$ and $\Omega^F r$. This choice is the same as
in~(3)
$\displaystyle u_z=-\frac{B_\phi B_z}{B_\phi^2+B_z^2}\Omega^F r,
\quad u_\phi=-\frac{B_z^2}{B_\phi^2+B_z^2}\Omega^F r$
for the value of $K$ equals to the $-\Omega^F r B_\phi/B^2$.
It can be checked that under change of $\phi$ to $-\phi$ the value of
$Q$ is not changed,
and the sign of $U$ is reversed.
Thus, as a result of integration $U=0$ and either $\chit_{res}=0$ or
$\chit_{res}=\pi/2$ as it is when we do not take into account Doppler boosting
and electric field.

We performed numerical calculations according to~(50) and~(51) using~(48)
and~(49).
\begin{figure}
\vspace{10cm}
\caption{Dependences of polarization $\Pi$ on $\Omega$~(a), on $v_{vis}$
for $\gamma=2.1$~(b) and for $\gamma=3.$~(c). Emitting particles are
concentrated close to the Alfv\'en resonant surface.
Here $B_\phi=+\Omega^F r B_z$, fundamental eigenmode with $m=1$.
Notations are the same as in Fig.~5. In Fig.~(a)
solid curves are for $\gamma=2.1$, dashed are for $\gamma=3.$}
\label{fig7}
\end{figure}
As above we take $B_z=\mbox{constant}$, $\Omega^F=\Omega(1-(r/R)^2)$,
$B_\phi=\pm \Omega^F r B_z$. Results for uniform particle distribution do
not match observations. Taking into account presence of the electric
field and Doppler boosting changes the polarization of synchrotron
emission essentially. Now only for small values of angle $\theta$ and
high~$\Omega$ apparent direction of the magnetic field as it is usually
derived from polarization measurements (just perpendicular to the electric
vector in the polarized component of the radiation) is transversal to the
jet axis. For most of inclination angles~$\theta$ and angular velocities
$\Omega$ apparent magnetic field is longitudinal. One can also see from
Fig.~7, where we plotted~$\Pi$
versus~$v_{vis}$ for $B_\phi=+\Omega^F rB_z$,
that our simulation does not match observational data. However, for
$B_\phi=-\Omega^F rB_z$ we are able to reproduce observations well enough,
if particles are distributed close to the Alfv\'en surface and $m=1$ or
$m=-1$. As for the case without Doppler boosting effect and without
taking account electric field, $\Pi$ is not changed when one changes
the sign of $\Omega$, but~$\Pi$ is changed when one put $B_\phi=
-\Omega^F rB_z$ instead of $B_\phi=\Omega^F rB_z$. Let us list
all transformations
which do not change equations for radial modes~(5) or lead to complex
conjugated equations for $\xi^\star$:
\begin{enumerate}
\item $\omega\to -\omega^\star$, $\,k\to -k$, $\,\Omega\to -\Omega$;
\item $\omega\to -\omega^\star$, $\,k\to -k$, $\,m\to -m$;
\item $\Omega\to -\Omega$, $\,m\to -m$;
\item $k\to -k$, $\,B_\phi=+\Omega^F rB_z\, \to\, B_\phi=-\Omega^F rB_z$;
\item $B_z\to -B_z$.
\end{enumerate}
So, if one set of parameters satisfies equations~(5) than parameters with
changed signs will also satisfy equations~(5) for the same~$\xi$ or
for~$\xi^\star$. Obviously, these substitutions are valid for $k_{min}$ and
$\omega_{min}$ as well. The value of~$r_A$ also is not changed under
transformations listed above. Degree of polarization~$\Pi$ is not influenced
by changing sign of~$B_z$ as well as by changing sign of~$\Omega$.
The crucial fact is that changing sign in relation $B_\phi=\pm \Omega^F rB_z$
leads to changing sign of $v_{phas}=\omega_{min}/k_{min}$, whilst all other
four transformations change neither~$\Pi$ nor they change~$v_{phas}$.
Therefore, in the sense of correlation between~$\Pi$ and $v_{vis}$, which
can be verified experimentally, all different cases are reduced to considering
both signs in the relation $B_\phi=\pm \Omega^F rB_z$, but we can restrict
ourself to $B_z>0$, $m>0$ and all $\Omega$ for which $r_A$ is located
inside the jet. Using formulae~(48), (49), (50) and expressions (3) for $u_z$
and $u_\phi$ one can check that
Stocks parameters remain unchanged when one changes sign of $B_\phi$ and
replace angle $\theta$ by $\pi-\theta$. Also, simultaneous changing of
sign of $v_{phas}$ and $\theta$ to $\pi-\theta$ leads to that $v_{vis}$
is transformed into $-v_{vis}$. Therefore, dependences of $\Pi$ on $v_{vis}$
in the case $B_\phi=-\Omega^F rB_z$ can be obtained from same dependences
in the case $B_\phi=+\Omega^F rB_z$ by reflection $v_{vis}\to -v_{vis}$
and relabelling each curve from $\theta$ to $\pi-\theta$.

We plotted degree of linear polarization~$\Pi$ versus~$v_{vis}$ for $m=1$ in
Figs.~7 and~8 for $B_\phi=+\Omega^F rB_z$
and $B_\phi=-\Omega^F rB_z$
correspondently. Alfv\'en resonance exists for $0.7<\Omega<4.9$. Therefore,
along curves in Figs.~7 and~8 the value
of $\Omega$ changes in this
interval. We see that curves in Fig.~7 do not match
observational data for
knots in jets, while those in Fig.~8 do. Dependence on the power law
index~$\gamma$ of emitting particles is small. We plotted results for $\gamma
=2.1$ and $\gamma=3.$ which are maximum and minimum values of $\gamma$
obtained in the process of the formation of the spectrum (see section~3).

\begin{figure}
\vspace{10cm}
\caption{Dependences of polarization $\Pi$ on $\Omega$~(a), on $v_{vis}$
for $\gamma=2.1$~(b) and for $\gamma=3.$~(c). Emitting particles are
concentrated close to the Alfv\'en resonant surface.
$B_\phi=-\Omega^F r B_z$, fundamental eigenmode with $m=1$
Notations are the same as in Fig.~5. In Fig.~(a)
solid curves are for $\gamma=2.1$, dashed are for $\gamma=3.$}
\label{fig8}
\end{figure}

We can suggest the following explanation of the fact that knots
in jets with $B_\phi=
-\Omega^F rB_z$ are observed more often than in jets having the opposite
winding of the magnetic field. If we look at the models of the formation
of magnetically driven jets from accretion discs (see, e.g.~Pelletier~\&
Pudritz, 1992) we
will find that magnetic field lines are always bent in the direction opposite
to the direction of disc rotation as soon as the velocity of the outflow is
directed outwards from the disc plane. This means that in the jet
approaching the observer $B_\phi=-\Omega^F rB_z$, whilst in the jet receding
from the observer $B_\phi=+\Omega^F rB_z$. Disturbances propagating along
approaching jet should be brighter than those in the counter-jet due to
the Doppler boosting effect. Therefore, we will observe knots in parts
of jets having $B_\phi=-\Omega^F rB_z$ more often than in those with
$B_\phi=+\Omega^F rB_z$. This explains the fact that the number of sources
having $\Pi>0$ is approximately the same as the number of sources having
$\Pi<0$ despite that for most of the values of parameters calculations give
longitudinal polarization $\Pi<0$. Also the difference between jets in
BL~Lac objects and jets in quasars finds its explanation by systematically
different values of~$\Omega^F$. In the frame of our model jets in BL~Lac
objects have higher value of~$\Omega^F$ than jets in quasars. Although we
performed calculations only for one particular profile of $\Omega^F(r)$
which seems to be close to what one should obtain from theory of the
formation of jets~(Istomin \& Pariev, 1996) we found this
tendency and it should be general
for profiles of $\Omega^F(r)$ which are qualitively similar to chosen one.
There are many physical arguments though~(Istomin~\& Pariev, 1994;
Istomin~\& Pariev, 1996; Beskin, Istomin~\& Pariev, 1992a),
why the real dependence
$\Omega^F(r)$ should have qualitative properties similar to
$\Omega (1-(r/R)^2)$ that we chose in our computations.

\section{Summary}

In present work we considered possible acceleration of electrons and
positrons inside relativistic rotating
electron-positron force--free cylindrical jet with spiral
magnetic field. It is very
plausible that inner parsec--scale  jets in active galaxies can be described
in the frame of this physical model.
 The observations of jets in 3C273 and M87 points out that the extragalactic
 jets are most probably electron-positron rather than electron-proton
 ( Morrison \& Sadun, 1992; Reynolds et al., 1996 )
Also it seems to be very likely that
such jets are not in steady state and contain waves excited by environmental
processes, of which most powerful would be the variability of accretion
rate onto central black hole and accretion disc instabilities both
strongly perturbing magnetic field at the base of the jet. We considered
the behaviour of such excitations inside cylindrical force--free jet
in our previous works (Istomin \& Pariev, 1994; Istomin \& Pariev, 1996)
and found that in wide range of parameters determining the equilibrium
structure of electromagnetic fields and perturbations inside the jet
there exist resonant surfaces on which the phase speed of eigenmodes is
equal to local Alfv\'en velocity. Those eigenmodes for which there is an
Alfv\'en resonance occured to have small damping decrement, whilst those
having no Alfv\'en resonances are neutrally stable. This fact points that
existence of Alfv\'en resonance leads to losses of energy of corresponding
eigenmodes.
 In present work we found how particles are accelerated in the
 strong fields near the Alfv\'en surface.
Acceleration process and
synchrotron losses combined together form power law energy spectrum of
ultra-relativistic electrons and positrons with index between 2 and 3 depending
upon initial energy of injected particles.
 The power law spectrum goes up to
 the energy $\e_{max}$, where the sharp fall down occurs (see Figs.~2,3).
 The magnitude of $\e_{max}$ depends on the initial particle temperature
 $T_0$ as well as on the characteristic acceleration energy $\e_1$(26),
 $\e_{max}\approx \e_1/\ln(\e_1/T_0)$. The quantity of $T_0$ evaluated
 from the equipartition condition in the acceleration region is equal to 
 $10$~MeV  for the quantities of magnetic field and density $10^{-2}\mbox{G}$
 and $0.1\mbox{cm}^{-3}$ correspondingly. As to the quantity of $\e_1$,
 it is, in accordance with (26), of the order of $10^4(\delta B/B_0)^2
 \mbox{MeV}$ for the same value of $B_0$, where $\delta B/B_0$ is the
 dimensionless amplitude of the perturbation. For large perturbations
 ($\delta B\approx B_0$) $\e_1$ reaches the values of the order of $10^3$MeV.
 Thus particles accelerated near Alfv\'en resonance are in the energy range
 of $10\mbox{MeV}<\e<10^3\mbox{MeV}$. These particles emitt synchrotron
radiation in the range of frequences approximately from  $1\mbox{MHz}$ 
to $100\mbox{GHz}$ which covers the frequences of modern radio
observations.

In order to obtain observable predictions of our theory of standing
wave excitations inside jets, we computed synchrotron radiation emitted
by accelerated particles. We considered two possibilities:
1)~ultra-relativistic
particles distributed uniformly across the jet, 2)~ultra-relativistic particles
concentrated close to the Alfv\'en resonance surface. The latter distribution
follows from our acceleration theory and we compared it with the former one
which might be produced by other acceleration processes. We assumed uniform
magnetic field in the jet, so the equilibrium configuration of the jet is
determined only by the radial profile of angular rotational
velocity $\Omega^F(r)$
of magnetic field lines. In our computations we also specified
component of the velocity parallel to the magnetic field lines in a way to
minimize the kinetic energy associated with the flow and took $\Omega^F(r)=
\Omega(1-(r/R_{jet})^2)$. The actual velocity is determined by the processes
of jet formation which is poorly understood now, still such minimal velocity
is physically favoured~(Novikov \& Frolov, 1986). The prominent characteristic
of synchrotron radiation is its polarization: degree of linear polarization
and orientation of electric vector with respect to the jet axis. We
calculated these quantities for the emission integrated over the jet cross
section taking into account relativistic bulk spiral flow of plasma
in the jet, which leads to the presence of electric field comparable by
magnitude with the magnetic field, Doppler boosting effect for the intensity
of synchrotron radiation and swing of the polarization position angle.
We want to stress that the last effect, which was first pointed out for the
simpler case of uniform magnetic field parallel to the jet axis by
Blandford \& K\"onigl in 1979, substantially modifies resulting degree of
linear polarization from the whole jet. However, because of the axial symmetry
of the problem, only two orientations of polarization position angle are
possible: parallel or perpendicular with respect to the jet axis.
Observations of polarization of VLBI scale jets (Gabuzda et al., 1994a)
also show bimodal distribution of the position angle: most of knots in
BL~Lac objects and quasars have polarization either close to the jet axis
or perpendicular to it.

We also calculated velocities of crests of standing eigenmodes and
compared them with the observed proper motions of bright VLBI knots in jets.
We were able to match observational data only for the case when emitting
particles are concentrated close to the Alfv\'en resonance surface and
magnetic field lines are twisted in the direction opposite to the
direction of rotation $\Omega^F$. The last fact can be understood by selection
effect due to the Doppler boosting which makes sources with jets approaching
the observer more favourable observed than receding jets.
Comparing  calculations with observations we can estimate angular
rotational velocity of jets. We obtained that $\Omega^F$ in BL~Lac objects
is intrinsically larger than in quasars which accounts for the difference in
polarization properties of these two classes of objects
(Gabuzda et al., 1994a): in BL~Lac the magnetic field inferred from
polarization position angle is oriented predominantly perpendicular to
the jet axis, in quasars it is parallel to the jet axis.

\begin{center}
{\Large \bf Acknowledgments}
\end{center}

Authors are grateful to Gabuzda~D.C. for compiling observational data
on polarization of quasars and helpful discussions of different
observational issues.
 This work was done under the partial support of the Russian Foundation
 for Fundamental Research
 (grant number 96-02-18203)
VIP acknowledges also partial support from European Southern Observatory
within the C\& EE Programme.

\newpage

\begin{center}
{\Large References}
\end{center}
\parindent0pt

Appl~S., Camenzind~M., 1992, A\&A, 256, 354

B\aa\aa~th~L.B., Zhang~F.J., 1991, A\&A, 243, 328

Begelman~M.C., Blanford~R.D. \& Rees~M.J. 1984, Rev. Mod. Phys., 56, 255 

Beskin~V.S., Istomin~Ya.N., Pariev~V.I., 1992a, in the book
Extragalactic Radio Sources~---~From Beams to Jets, edited by J.~Roland,
H.~Sol, G.~Pelletier. Cambridge University Press, Cambridge 

Beskin~V.S., Istomin~Ya.N., Pariev~V.I.,
1992b, AZh, 69, 1258 

Blandford~R.D., K\"onigl~A., 1979, ApJ, 232, 34

Blanford~R.D., Pringle~J.E., 1976, MNRAS, 176, 443 

Brown~L.F., Roberts~D.H., Wardle~J.F.C., 1994, ApJ, 437, 108

Gabuzda~D.C., Cawthorne~T.V., 1993, in Davis~R.J., Booth~R.S., eds,
Subarcsecond Radio Astronomy. Cambridge Univ. Press, Cambridge, p.211

Gabuzda~D.C., Cawthorne~T.V., 1996, MNRAS, 283, 759

Gabuzda~D.C., Cawthorne~T.V., Roberts~D.H., Wardle~J.F.C., 1989, ApJ, 347,
701

Gabuzda~D.C., Mullan~C.M., Cawthorne~T.V., Wardle~J.F.C.,
Roberts~D.H., 1994a, ApJ,~435,~140 

Gabuzda~D.C., Wardle~J.F.C., Roberts~D.H., Aller~M.F., Aller~H.D., 1994b,
ApJ, 435, 128

Ginzburg~V.L., 1989, Applications of Electrodynamics in Theoretical Physics and
Astrophysics. Gordon and Breach Science Publishers, New York, p.65 

Istomin~Ya.N., Pariev~V.I., 1994, MNRAS, 267, 629

Istomin~Ya.N., Pariev~V.I., 1996, MNRAS, 281, 1 

Mikhaylovskii~A.B., 1975, in Reviews of Plasma Physics, ed. by Leontovich H.A.,
Consultants Bureau, New York, Vol.6, p.77 

Morrison~P., Sadun~A., 1992, MNRAS, 254, 488.

Mushotzky~R.F., Done~C., Pound~K., 1993, Ann. Rev. Astron. Astrophys.,
717 

Novikov~I.D., Frolov~V.P., 1986, The Physics of Black Holes. Nauka Press,
Moscow

Pelletier~G., Pudritz~R., 1992, ApJ, 394, 117 

Rees~M.J., 1984, Ann. Rev. Astron. Astrophys. ,471 

Reynolds~C.S., Fabian~A.C., Celotti~A., Rees~M.J., 1996, MNRAS, 283, 873.

Sivukhin~D.V., 1965, in Reviews of Plasma Physics, ed. by Leontovich H.A.,
Consultants Bureau, New York, Vol.1, p.1 

Torricelli-Ciamponi~G., Petrini~P., 1990, ApJ., 361, 32 

Witzel~A., Schalinski~C.J., Johnston~K.J., Biermann~P.L., Krichbaum~T.P.,
Hummel~C.A., Eckart~A., 1988, A\&A, 206, 245

Witzel~A., Wagner~S., Wegner~R., Steffen~W., Kirchbaum~T., 1993, in
Davis R.J. \& Booth R.S., eds Sub-arcsecond Radio Astronomy.
Cambridge Univ. Press, Cambridge, p. 159

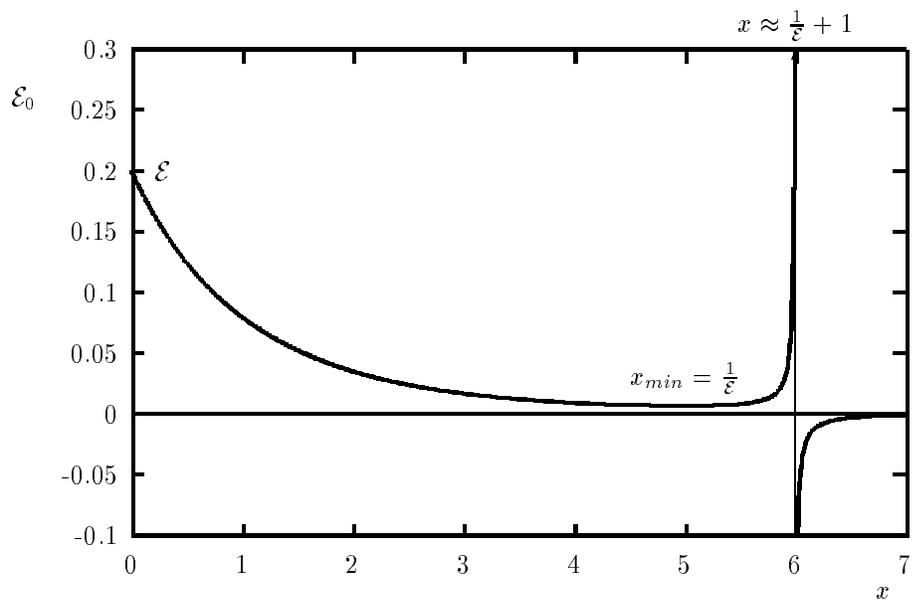
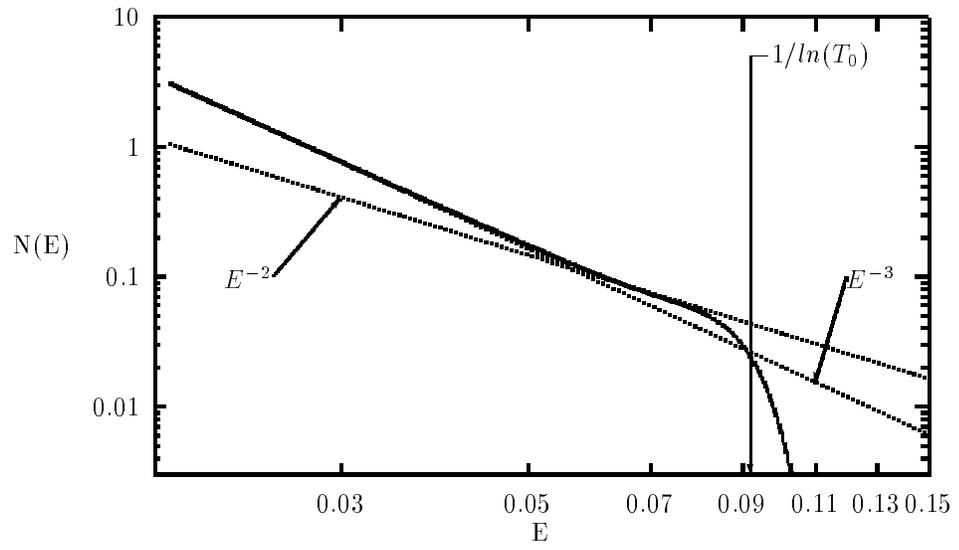
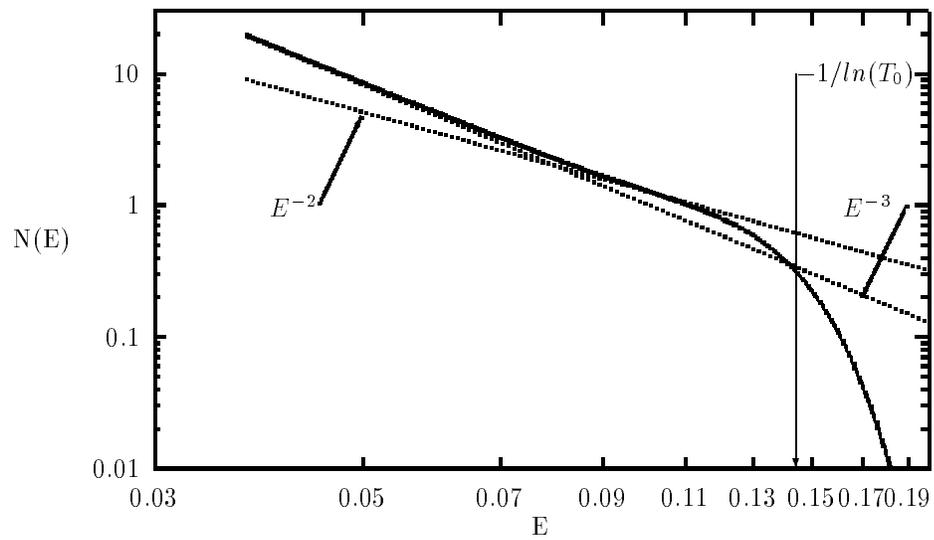
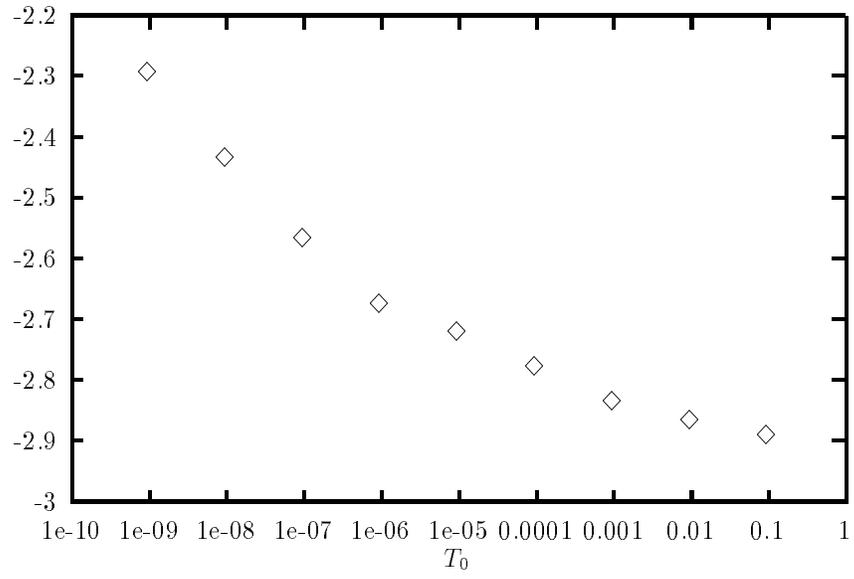
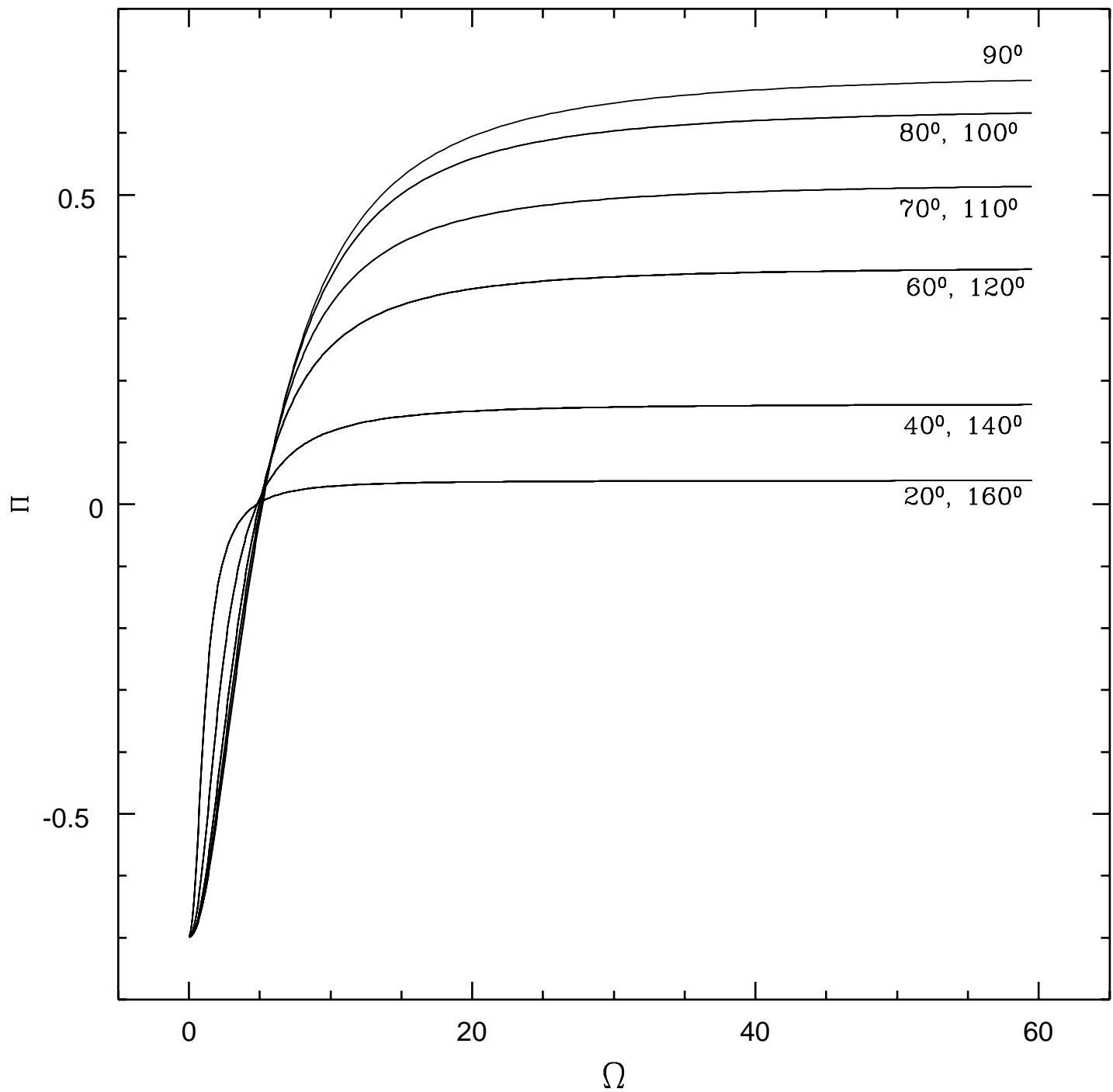
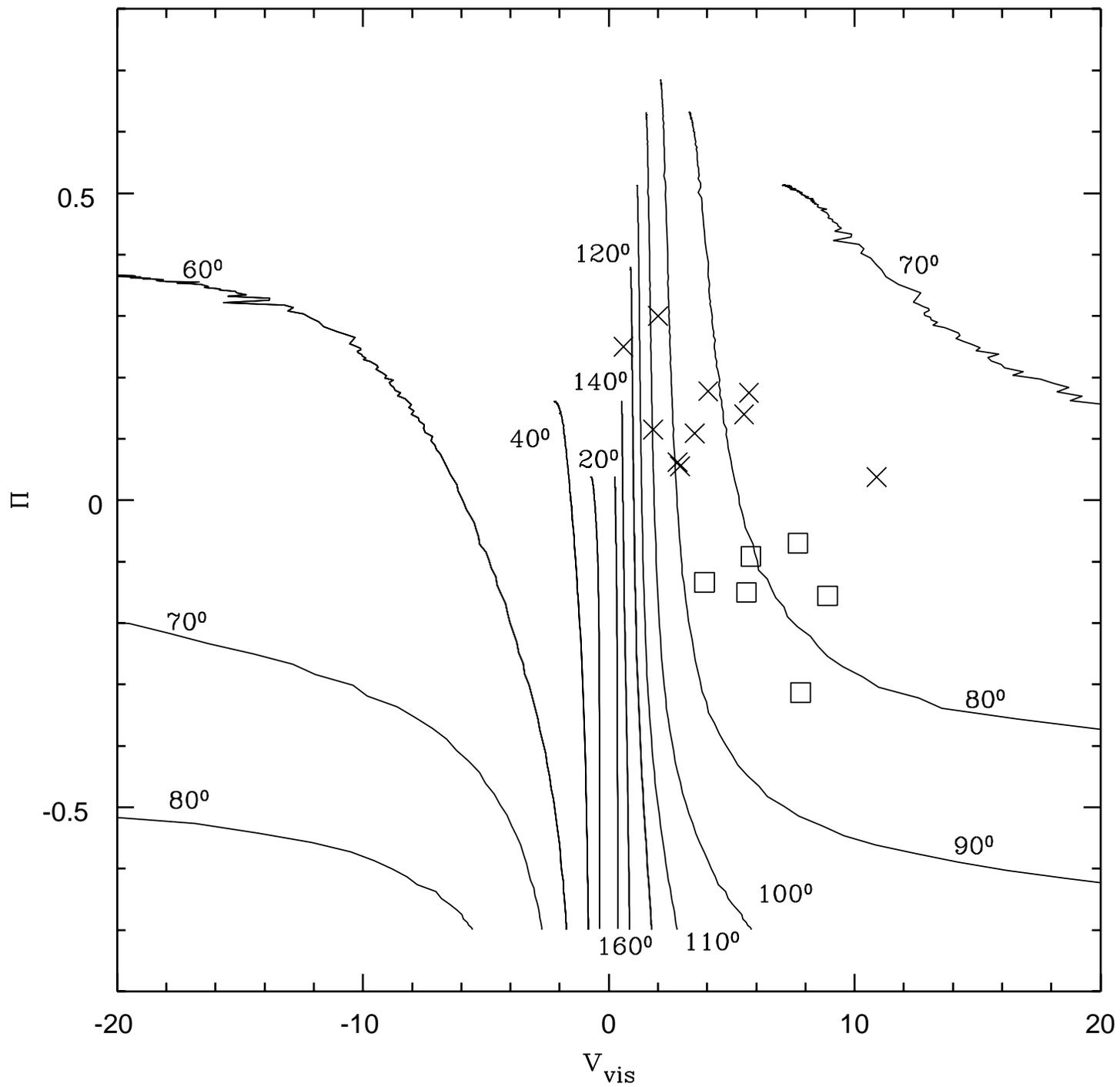
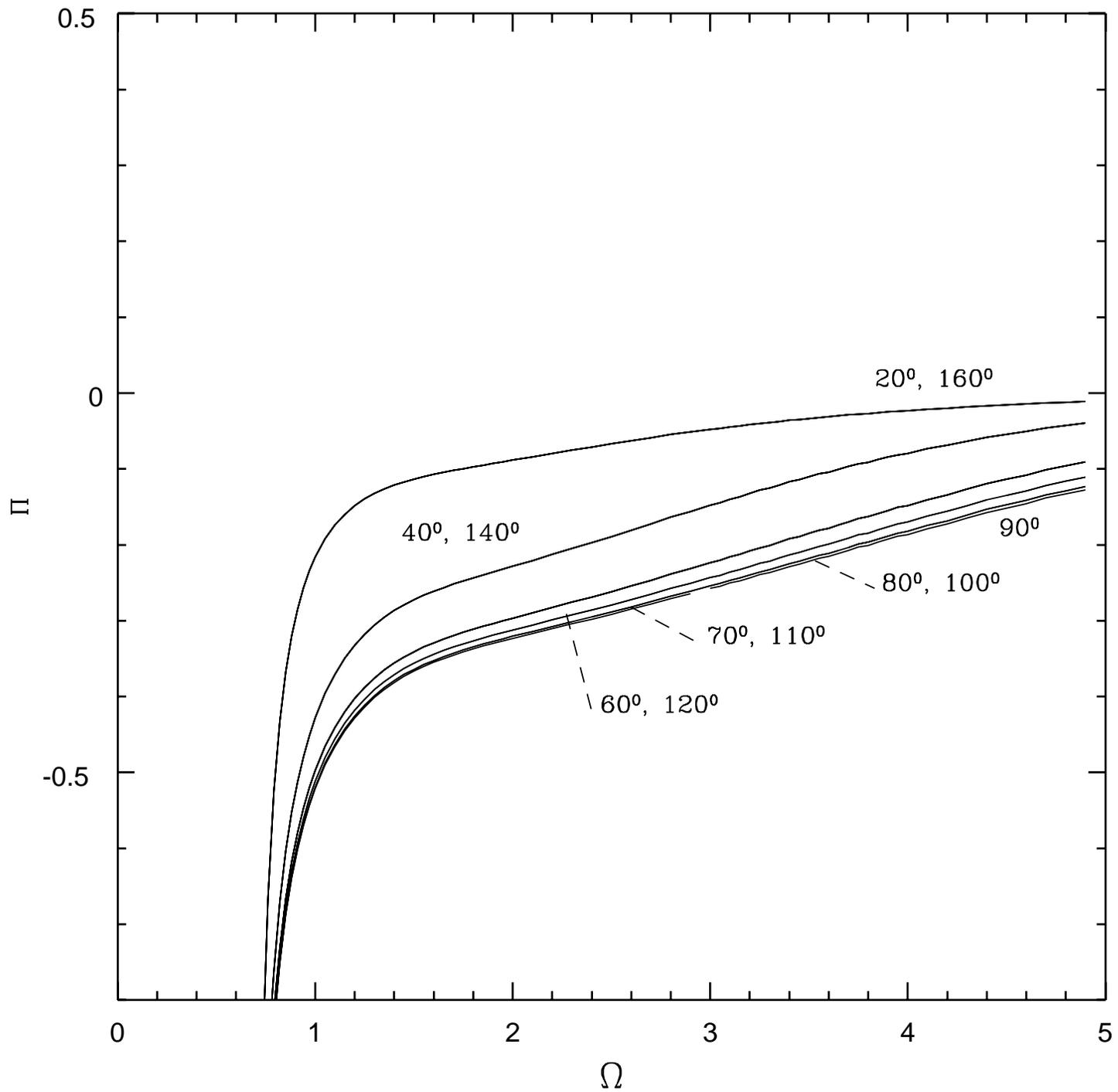
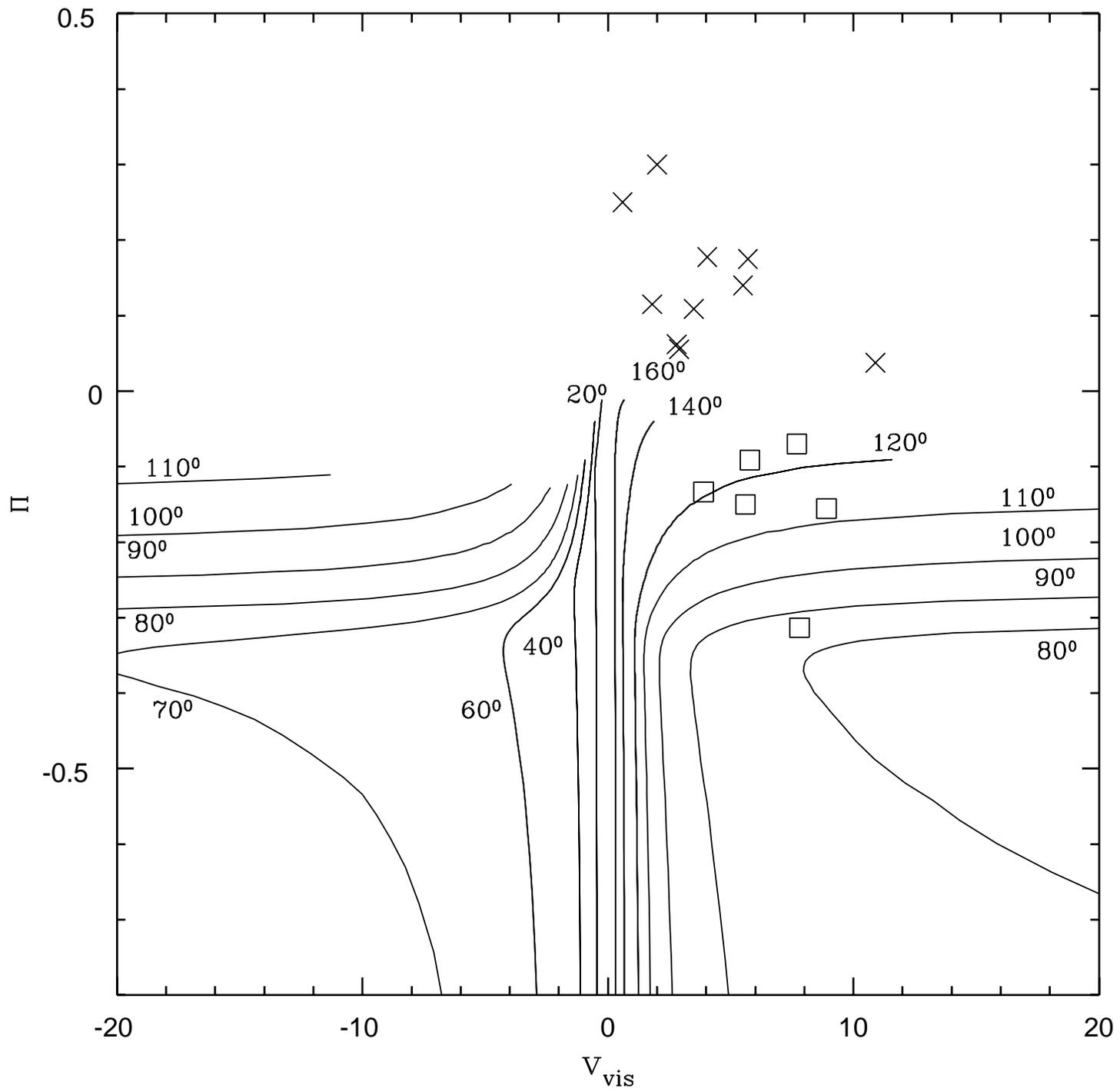
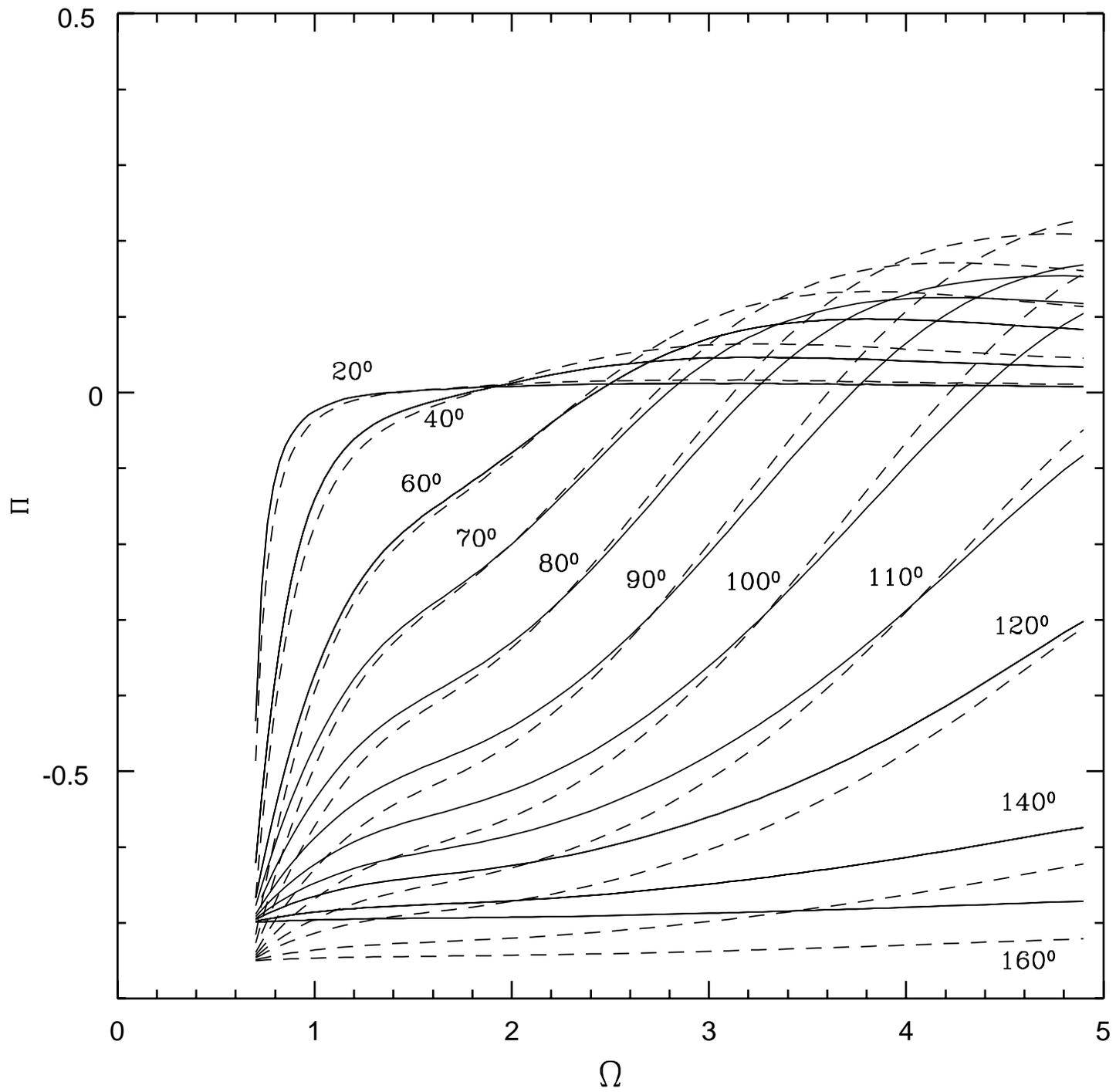
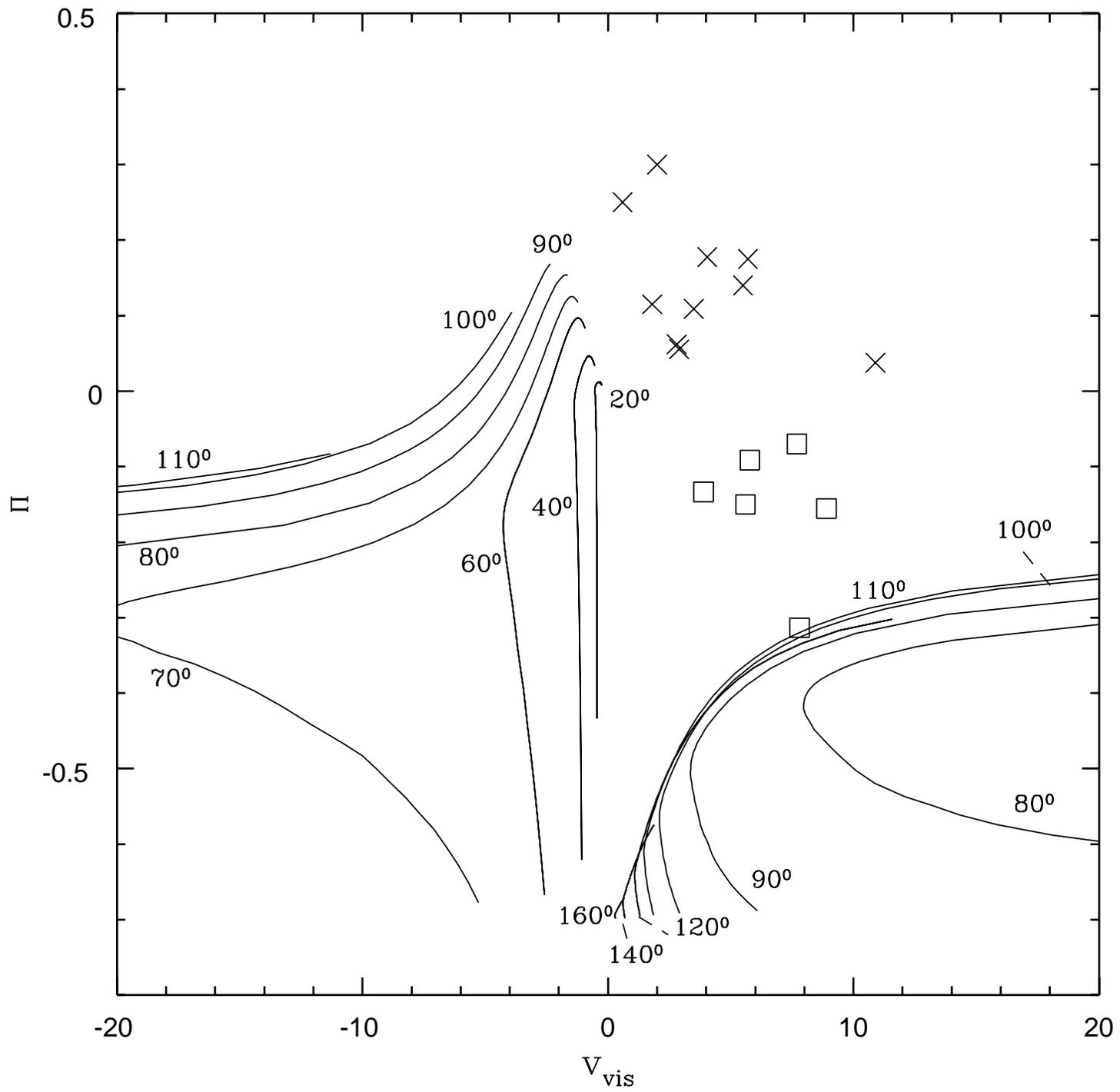
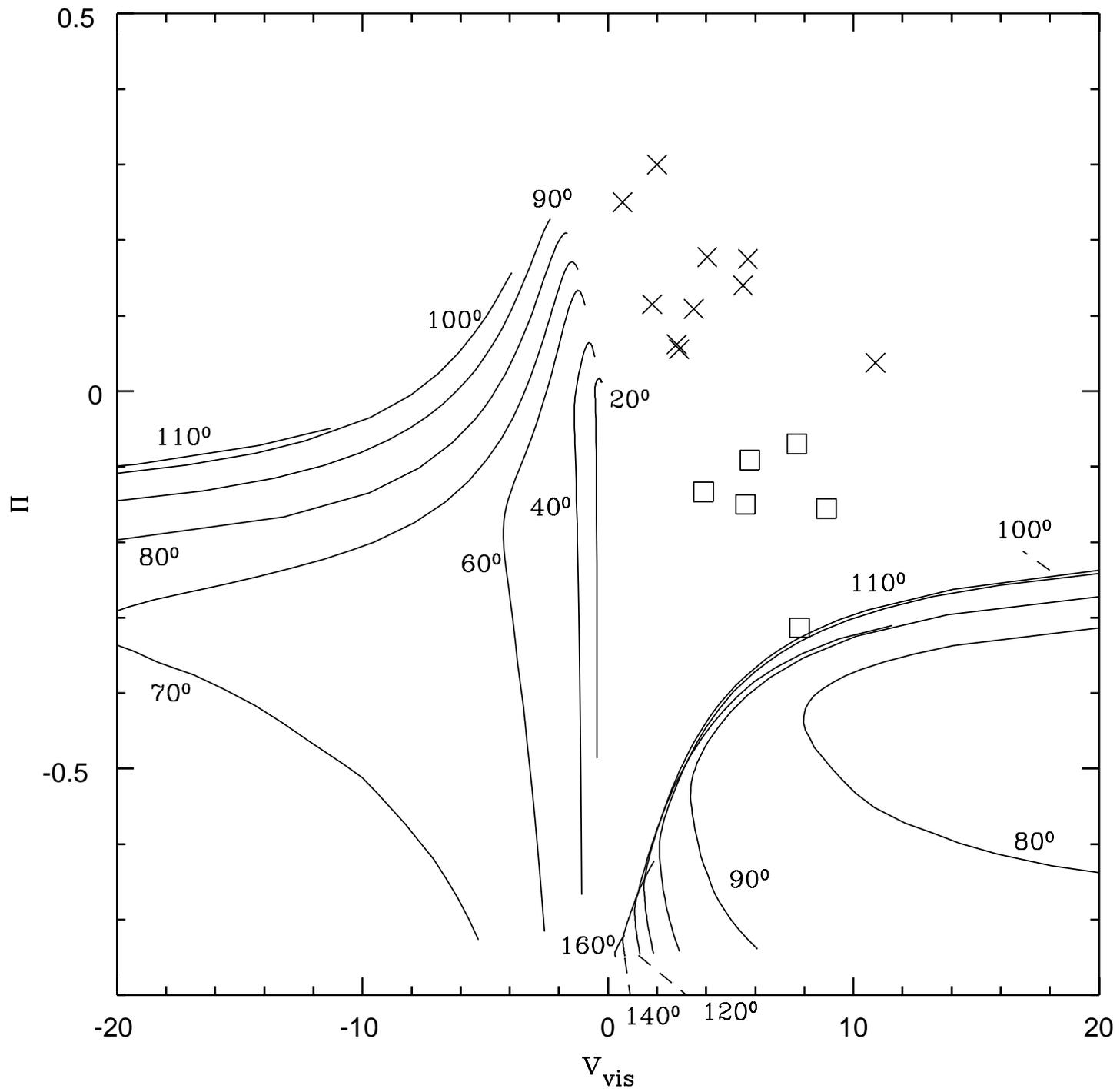
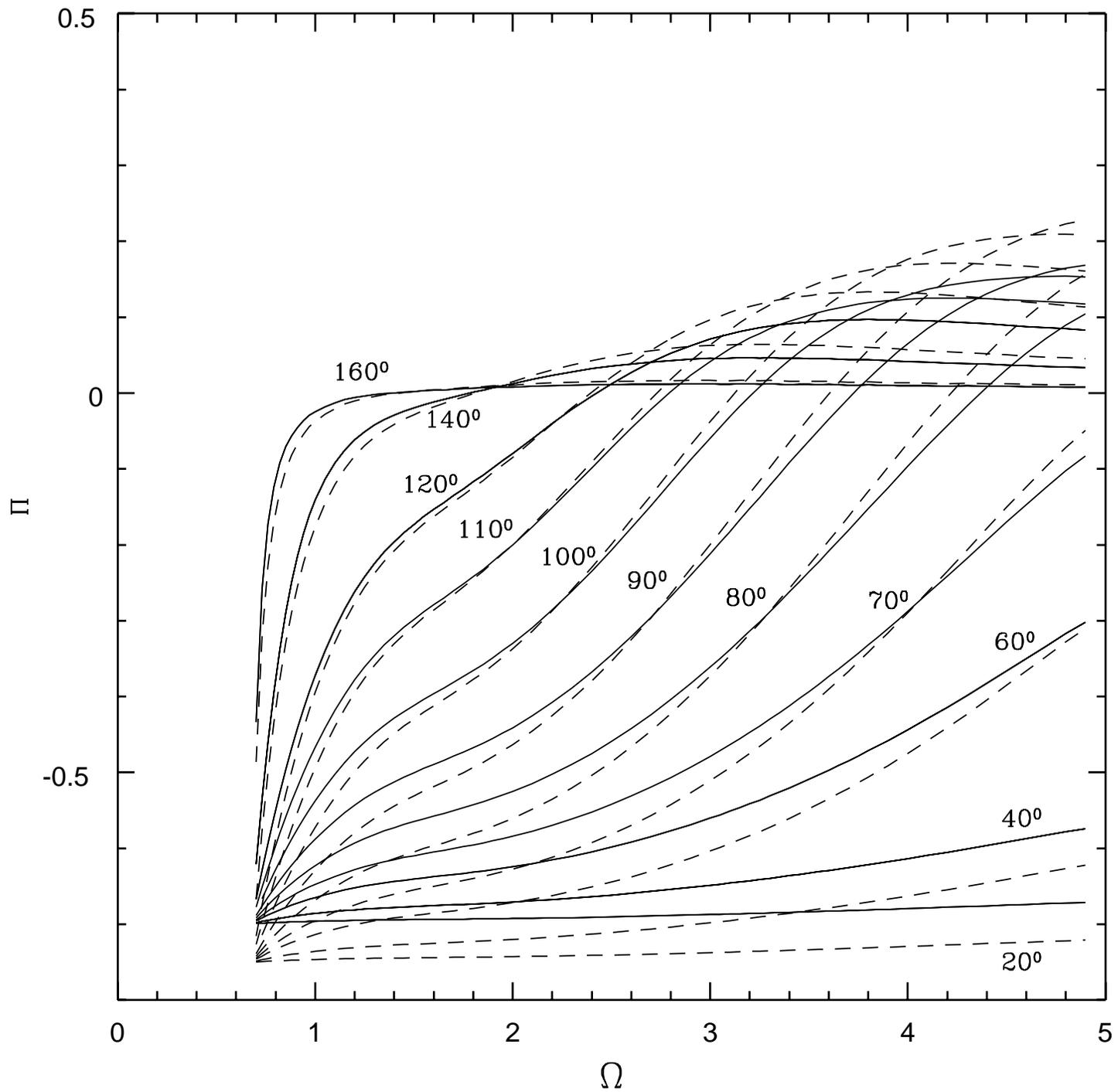
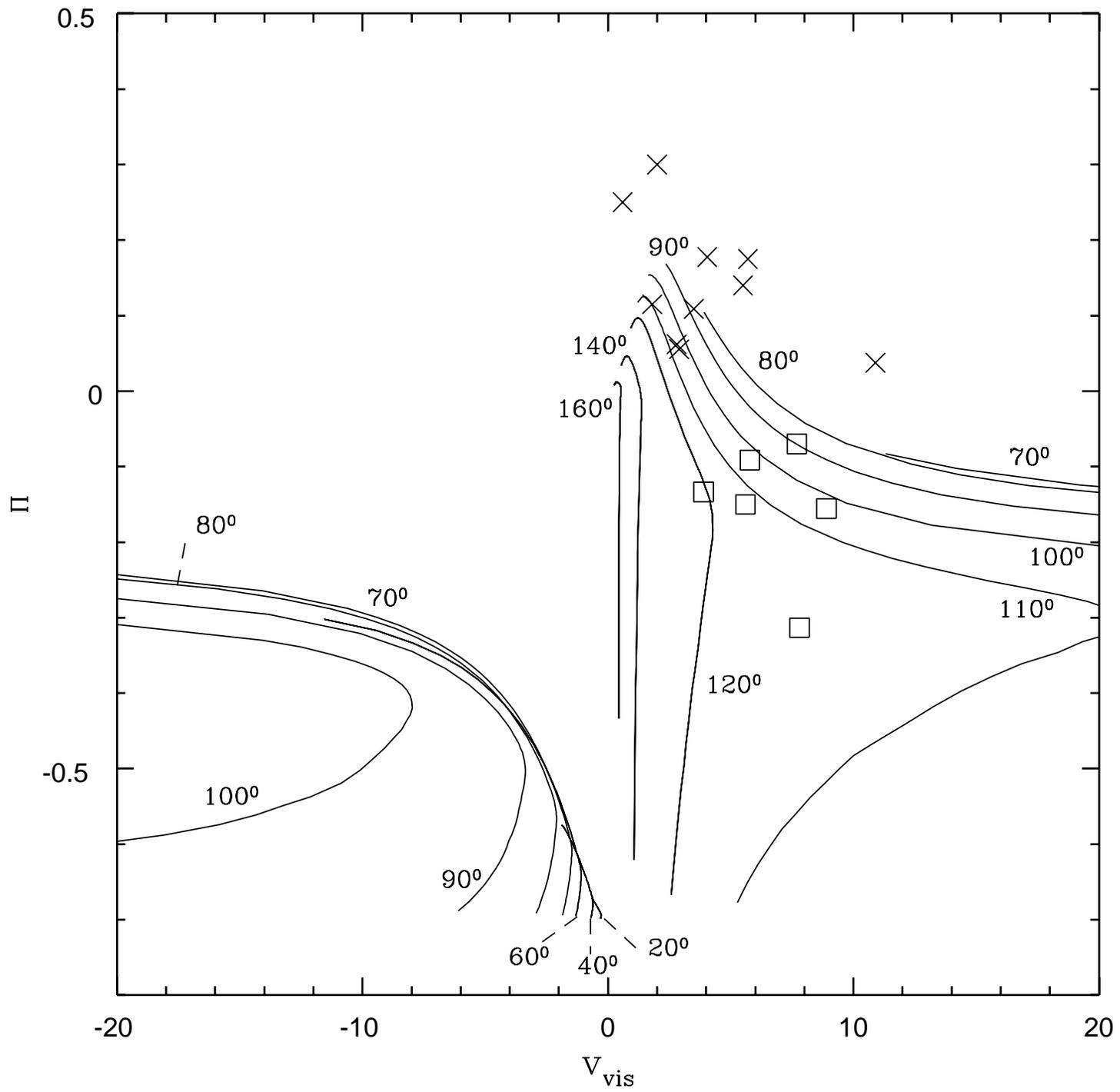
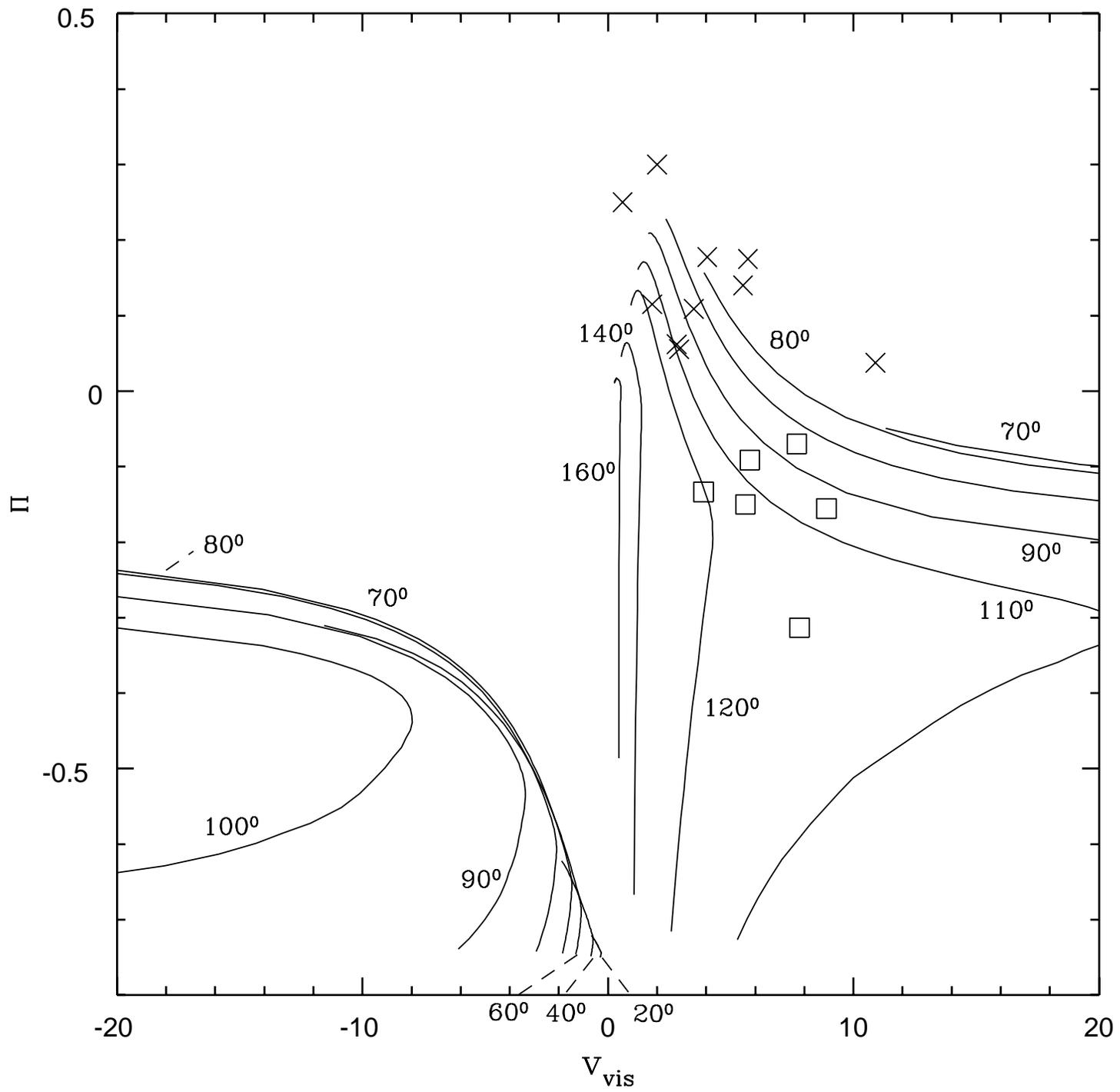
\end{document}